\documentclass[fleqn,usenatbib,useAMS]{mnras}

\usepackage[T1]{fontenc}
\usepackage{ae,aecompl}

\usepackage{graphicx}	
\usepackage{amsmath}	
\usepackage{amssymb}	
\usepackage{amstext}
\usepackage{amsfonts}
\usepackage{float}

\newcommand{\kms}{km\,s$^{-1}$}

\newcommand{\msun}{$M_\odot$}

\newcommand{\mbh}{M$_\bullet$}
\newcommand{\ml}{$\Upsilon$}
\newcommand{\rin}{\ensuremath{r_\textrm{infl}}}

\newcommand{\mmw}{M$_\mathrm{MWNSC}$}
\newcommand{\reff}{\ensuremath{r_\textrm{eff}}}
\newcommand{\chisq}{$\chi^2$}
\newcommand{\chired}{$\chi^2_{\mathrm{red}}$}

\newcommand{\nmod}{4899}
\newcommand{\bfchisq}{290}
\newcommand{\bfchired}{0.42}

\newcommand{\bfbh}{$3.0$}
\newcommand{\bfml}{0.90}
\newcommand{\bfq}{0.28}
\newcommand{\bfp}{0.64}
\newcommand{\bfu}{0.99}

\newcommand{\tentosix}{$\times10^6$}
\newcommand{\bhos}{$^{+1.1}_{-1.3}$}
\newcommand{\mlos}{$^{+0.76}_{-0.08}$}
\newcommand{\qos}{$^{+0.0}_{-0.02}$}
\newcommand{\pos}{$^{+0.18}_{-0.06}$}
\newcommand{\uos}{$^{+0.0}_{-0.01}$}
\newcommand{\bhts}{$^{+2.4}_{-2.3}$}
\newcommand{\mlts}{$^{+1.12}_{-0.32}$}
\newcommand{\qts}{$^{+0.0}_{-0.06}$}
\newcommand{\pts}{$^{+0.30}_{-0.22}$}
\newcommand{\uts}{$^{+0.0}_{-0.05}$}

\title[Triaxial orbit-based modelling of the Milky Way Nuclear Star Cluster]{Triaxial orbit-based modelling of the Milky Way Nuclear Star Cluster \thanks{Based on observations collected at the European Organisation for Astronomical Research in the Southern Hemisphere, Chile (289.B-5010(A)).}}

\author[A.~Feldmeier-Krause et al.]{A.~Feldmeier-Krause,$^{1,}$$^{2}$\thanks{E-mail:
afeldmei@uchicago.edu} 
L.~Zhu$^{3}$, N.~Neumayer$^{3}$, G.~van de Ven$^{3}$, P.~T.~de~Zeeuw,$^{1,}$$^{4}$
\newauthor
   R.~Sch{\"o}del$^{5}$ 
\\
$^{1}$European Southern Observatory (ESO), Karl-Schwarzschild-Stra{\ss}e 2, 85748 Garching, Germany\\
$^{2}$The University of Chicago, The Department of Astronomy and Astrophysics, 5640 S. Ellis Ave, Chicago, IL 60637, USA\\
$^{3}$Max-Planck-Institut f{\"u}r Astronomie, K{\"o}nigsstuhl 17, 69117 Heidelberg, Germany\\
$^{4}$Sterrewacht Leiden, Leiden University, Postbus 9513, 2300 RA Leiden, The Netherlands\\
$^{5}$Instituto de Astrof\'{i}sica de Andaluc\'{i}a (CSIC), Glorieta de la Astronom\'{i}a s/n, 18008 Granada, Spain
}

\date{Accepted 2016 December 21. Received 2016 December 19; in original form 2016 July 22}
\pubyear{2016}

\begin{document}
\label{firstpage}
\pagerange{\pageref{firstpage}--\pageref{lastpage}}
\maketitle

\begin{abstract}
We construct  triaxial dynamical models for the Milky Way  nuclear star cluster using  Schwarz\-schild's orbit superposition technique. We fit the stellar
kinematic maps presented in \cite{isaacanja}. The models are used to constrain the  supermassive black hole mass \mbh, dynamical mass-to-light ratio \ml, and the intrinsic shape of the cluster. Our best-fitting model has \mbh\space= (\bfbh\bhos)\,\tentosix\,\msun, \ml\space= (\bfml \mlos)\,\msun/$L_{\odot, 4.5\mu m}$, and a compression  of the cluster along the line-of-sight. Our results are in agreement with the direct measurement of the supermassive black hole mass using the motion of  stars on Keplerian orbits. 
 The mass-to-light ratio is consistent  with stellar population studies of other galaxies in the mid-infrared. It is possible that we underestimate \mbh\space and  overestimate  the cluster's triaxiality due to observational effects. The spatially semi-resolved kinematic data  and extinction within the nuclear star cluster  bias the observations to the near side of the cluster, and may appear as a compression of the nuclear star cluster along the line-of-sight.
 We derive  a total  dynamical mass for the Milky Way nuclear star cluster of  \mmw\space=  (2.1$\pm0.7$)\,$\times$\,10$^7$\,\msun\space within a sphere with  radius  $r$ = 2 $\times$ $\reff$\space = 8.4\,pc.
 The best-fitting model is tangentially anisotropic in the central $r$ = 0.5-2\,pc of the nuclear star cluster, but  close to isotropic at larger radii.  Our triaxial models are  able to recover complex kinematic substructures  in the velocity map.  
\end{abstract}




\section{Introduction}

The Milky Way nuclear star cluster is the ideal object to study the dynamics of a stellar system around a supermassive black hole. At a distance of 8\,kpc it  is close enough to resolve the individual  stars, and measure discrete velocities in three dimensions. Modelling the stellar kinematics can constrain  the mass distribution of the star cluster, and reveal the presence of a central dark massive object. In the special case of our own Galaxy, it it possible to observe Keplerian orbits of stars around a dark, point-mass-like object in the Galactic centre.  These observations  constrain this dark object to be a supermassive black hole with a mass of (4.1 $\pm$ 0.6)\,\tentosix\,\msun\space   \citep{ghez08},  (4.3 $\pm$ 0.39)\,\tentosix\,\msun\space \citep{gillessen09}, or (4.02 $\pm$ 0.20)\,\tentosix\,\msun\space   \citep{2016ApJ...830...17B}. 
Unfortunately, similar  high-resolution observations are not yet possible in other galaxies. 

Already in the 1970s the requirement of a central supermassive black hole in the Galactic centre was discussed   to explain observational data  \citep[e.g.][]{1977ARA&A..15..295O}.  Several studies  used stellar radial velocities to  constrain  the  mass distribution in the Galactic centre \citep[e.g.][]{1988ApJ...330L..33R,mcginn89,sellgren90,haller96,genzel96}. 
 Also stellar proper motions were used to study the Galactic centre mass distribution  \citep{Rainerpm09}. 
  Several   studies  combined radial velocity and proper motion data \citep{trippe08,tuan,fritz16}. 
 The mass distribution was derived using the  spherical \cite{jeans} equations or  the  projected mass estimators of \cite{1981ApJ...244..805B} for spherical systems. 
 These studies found that a central dark mass of 2$-$5\,$\times$\,10$^6$\,\msun\space is required to explain  the observations.

 Together with the increase of observational data, also the modelling became more advanced. \cite{trippe08} included the rotation of the nuclear star cluster in the modelling, although the rotation velocity of their data  was too high \citep{Rainerpm09,isaacanja}. 
 \cite{isaacanja} and \cite{chatzopoulos} studied the Milky Way nuclear star cluster using axisymmetric Jeans models. \cite{chatzopoulos} showed the advantages of axisymmetric models over spherical Jeans models, which cannot explain the observed asymmetry of the velocity dispersion of proper motions parallel and perpendicular to the Galactic plane. 
 The nuclear star cluster appears to be flattened in its light distribution \citep{sb} as well as in  the kinematics \citep{chatzopoulos}. Most studies showed that the nuclear star cluster kinematics is in agreement with isotropy \citep{Rainerpm09,tuan,chatzopoulos}, although the uncertainties are quite large (e.g. \citealt{tuan}).  All these models assumed a constant mass-to-light ratio  for the light distribution of  the cluster.

In this study we relax the assumption of axisymmetry and use  triaxial orbit-based \cite{1979ApJ...232..236S} models. Orbit-based models make no assumptions on the velocity  anisotropy  of the  stellar motions, as Jeans models do. Further, the higher moments of the kinematics can also be included \citep{1997ApJ...488..702R}, which is important to break the degeneracy of mass and anisotropy in dynamical models. 

Orbit-based models are commonly used to analyse line-of-sight velocity data of other galaxies \citep[e.g.][]{1998ApJ...493..613V,2000AJ....119.1157G,2005ApJ...628..137V,remco08}, 
and are an excellent tool to detect and measure the masses of supermassive black holes and dark matter halos.                                           
For extragalactic systems, the data are usually obtained from integrated light observations. Each data point contains the accumulated kinematics of many stars, weighted by their respective brightness. However, modelling the dynamics of integrated light data may be prone to systematic uncertainties, and bias the results of the central black hole mass. Therefore, it is interesting to test dynamical models on systems for which we know the central black hole mass from other independent measurements. The Milky Way nuclear star cluster is a good object for this kind of test. Also megamaser disc galaxies are useful to validate stellar dynamical black hole measurements. Black hole  mass measurements from megamasers are very precise with uncertainties of only about 10 per cent. However, there is currently only one megamaser disc galaxy with a stellar dynamical black hole mass measurement  \citep{2016ApJ...819...11V}, NGC 4258. Different dynamical studies found either a 15 per cent lower or a 25 per cent higher  black hole mass than the maser measurement \citep{2009ApJ...693..946S,2015MNRAS.450..128D}.

We use the triaxial orbit-based  code  by \cite{remco08} to model the light distribution and line-of-sight kinematics of the Milky Way nuclear star cluster. We use the spectroscopic data cube constructed by \cite{isaacanja} for the  kinematic data, and derive a surface brightness distribution using \textit{Spitzer} 4.5\,$\micron$ and NACO $H-$band images. 
We assume a galactocentric distance of 8\,kpc \citep{distance8kpc} and a position angle 31\fdg40 East of North (J2000.0 coordinates, \citealt{2004reid}) with respect to the Galactic plane.
This paper is organised as follows: We describe the kinematic and photometric  data  in Section \ref{sec:datasw}. The dynamical models are introduced in Section \ref{sec:model}. Section \ref{sec:discsw} discusses the results, and Section \ref{sec:sumsw} summarizes the main conclusions.

\begin{figure*}
  \centering  
    \includegraphics[width=0.95\textwidth]{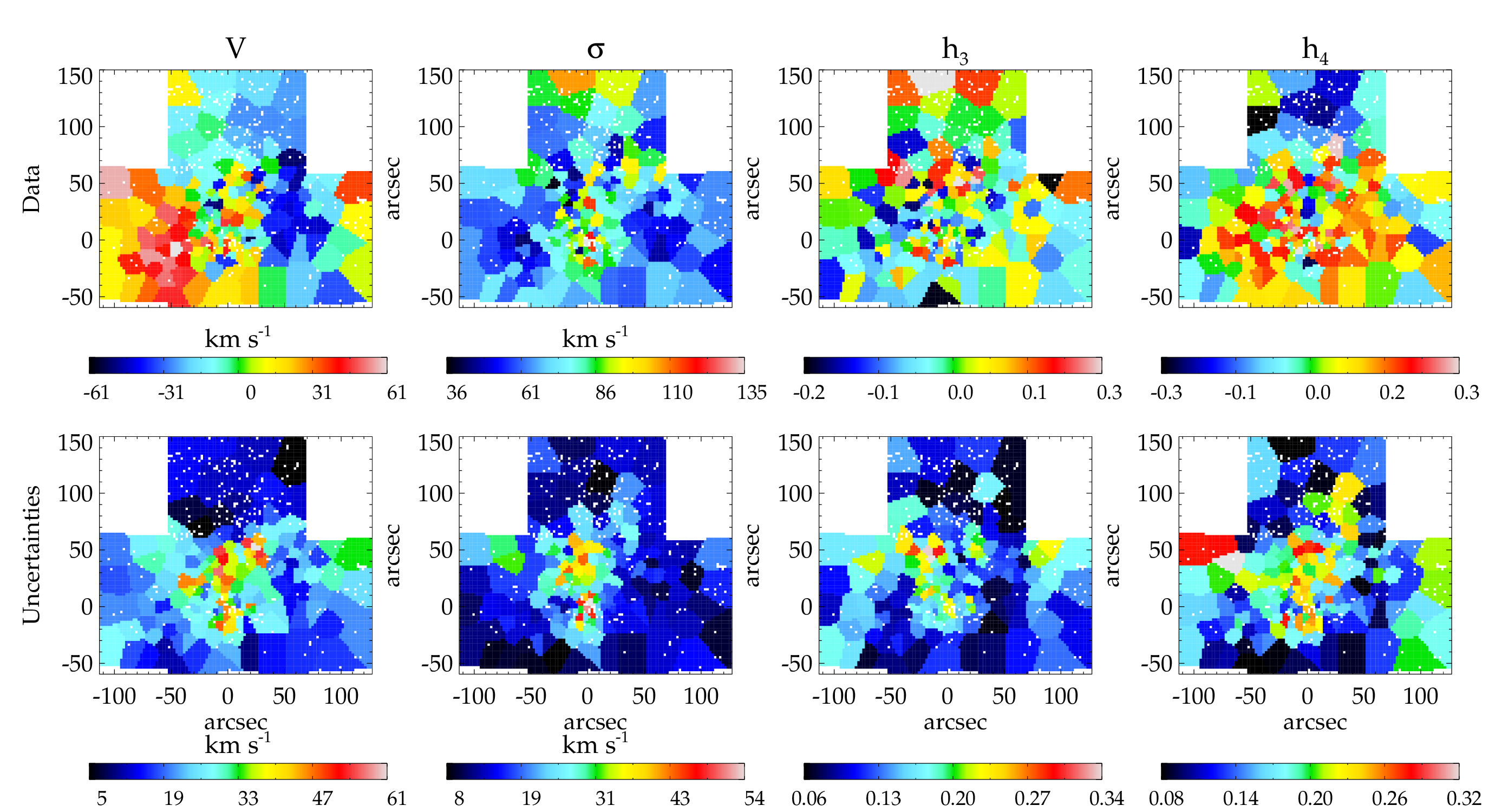}
 \caption{Kinematic data  (top row) and  respective uncertainties (bottom row). The columns denote  velocity $V$, velocity dispersion $\sigma$, Gauss-Hermite moments $h_3$, and $h_4$. White pixels are  due to excluded bright stars.  }
	\label{fig:swkindata}
\end{figure*}

\section{Description of the Data}
\label{sec:datasw}
\subsection{Kinematic data}
\label{sec:kin}
The line-of-sight velocity distribution (LOSVD) provides constraints on the dynamical structure of stellar systems. To extract this information, we used the near-infrared $K-$band spectroscopic data cube of \cite{isaacanja}, which has a pixel scale of 2.22\,arcsec pixel$^{-1}$.  We used the data cube that was cleaned of foreground stars and bright stars. The cleaned data cube contains the light of the old red giant star population. 

We fitted the LOSVD as in \cite{isaacanja}, i.e.  on the  stellar CO absorption lines  (2.2902$-$2.365\,$\micron$) with the IDL routine \textsc{pPXF} \citep{ppxf} and  the  high resolution spectra of \cite{wallace} as template stars.  We applied the same spatial binning as \cite{isaacanja}, resulting in 175 spatial bins. \cite{isaacanja} fitted only  the velocity $V$ and velocity dispersion $\sigma$.  However, we fitted in addition also higher  moments of the LOSVD,  in particular the Gauss-Hermite parameters $h_3$ and $h_4$.  
We added noise to each  of the 175 integrated light spectra  in 100 Monte Carlo simulation runs and  obtained a distribution for each moment of the LOSVD. The mean and standard deviation of the Monte Carlo distribution are taken as  measurement and 1$\sigma$ uncertainty of the kinematics.

Since the Milky Way nuclear star cluster is at a distance of only 8\,kpc, the spectroscopic observations are spatially semi-resolved. Bright stars can be resolved individually, and  contribute a large fraction of the flux. For that reason we used the cleaned data cube of \cite{isaacanja}, where bright stars were excluded. However, the kinematic maps still show stochastic shot noise. As a consequence, the difference of the kinematics in adjacent bins can be higher than their uncertainties, which causes problems when we model the kinematics. The stochastic noise can be mistaken for signal, and this means the best fit will be achieved by modelling the shot noise. To prevent this,  we increased our kinematic uncertainties $\epsilon_{V}$ such that the difference of the measurement in two adjacent bins ($V_i - V_j$) is less than the sum of their uncertainties ($\epsilon_{V_i} + \epsilon_{V_j}$).  We did this for the velocity $V$, velocity dispersion $\sigma$, $h_3$, and $h_4$, and find that it is required for  about 68 per cent of the kinematic data uncertainties. Additionally, we point-symmetrised the kinematics using the procedure of \cite{2010MNRAS.401.1770V}.  The median uncertainties or $V$, $\sigma$, $h_3$, and $h_4$ are  24.6\,\kms,  18.4\,\kms,     0.15, and    0.17. 
Our resulting kinematic maps are consistent with the maps of \cite{isaacanja}. We find rotation in the velocity map of approximately 50\,\kms\space and an increase in the velocity dispersion   from about 65\,\kms\space towards  $\sigma_{\textrm{max}}$=135\,\kms\space at the centre. The kinematic maps are shown on the top row of Fig. \ref{fig:swkindata}, the uncertainties are shown on the bottom row.

\subsection{Imaging data and surface brightness distribution}
The light distribution of the nuclear star cluster traces the stellar density. We require the  two-dimensional light distribution of the red giant stars, which are our kinematic tracers. 
The extinction is high at optical wavelengths  in the Galactic centre \citep[$A_V$\,$\sim$\,30\,mag,][]{scoville03,2013EP&S...65.1127G}, therefore we used near- and mid-infrared images.

For the central 40.4\,arcsec\,$\times$\,40.4\,arcsec (1.6\,pc\,$\times$\,1.6\,pc) we used the high-resolution NACO $H$-band mosaic of \cite{Rainerpm09}, which has a  spatial  scale of 0.027\,arcsec\,pixel$^{-1}$. We preferred the $H$ band over the $K$ band in order to avoid light  from   gas emission lines   in the $K$ band \citep[Br\,$\gamma$ and He\,I,][]{paumard04}.
Our kinematic tracers are cool late-type stars, but there are also more than 100 hot, young stars located in the centre of the cluster, within a projected radius $r$\,=\,0.5\,pc ($\sim$12.8\,arcsec, \citealt{paumard06}). We masked out the young stars from the image with a 15 pixel radius. For the bright red supergiant IRS~7 we used a larger mask with a 30 pixel radius.  Beyond the central 0.5\,pc, the nuclear star cluster light is dominated by cool stars, and the contribution of young stars is negligible \citep{kmoset}.   

For the large-scale light distribution, we used  \textit{Spitzer} IRAC images  \citep{iracim}. These images were corrected for dust extinction and PAH emission by \cite{sb}. We used the extinction and emission corrected 4.5\,$\micron$ image to measure the light distribution. The image was smoothed to a  scale of 5\,arcsec pixel$^{-1}$, and extends over $\sim$270\,pc\,$\times$\,200\,pc. We excluded a central circle with $r$\,=\,0.6\,pc ($\sim$15.4\,arcsec)  to avoid contribution from  ionised gas emission and  young stars. In addition we masked out the young Quintuplet star cluster \citep{figer99}, and the  dark 20-\kms-cloud M-0.13-0.08 \citep{20kmsref}.

We used  the \textsc{MGE\_FIT\_SECTORS} package \citep{mgeidl} to derive the surface brightness distribution.  The Multi-Gaussian-Expansion model \citep{mgeeric} has the advantage that it can be deprojected analytically. 
 We measured the photometry  of the two images along the major axis and the minor axis. We assumed that the cluster's major axis is aligned along the Galactic plane, as found by \cite{sb}, and constant. The centre  is the position of Sgr~A*, which is the radio source associated with the Galactic centre supermassive black hole. We fitted a scale factor to match the photometry of the two images in   the region where they overlap (16$-$27.8\,arcsec). Then  we measured the  photometry on each image along 12 angular  sectors,  and  converted the NACO photometry to the \textit{Spitzer} flux. 
 Assuming   four-fold symmetry, the measurements of four quadrants are averaged on elliptical annuli with constant ellipticity.  Using the  photometric measurements of the two images, we fitted   a set of two-dimensional Gaussian functions, taking the point-spread-function (PSF) of the NACO image into account. 
  
\begin{figure}
\begin{center}
\includegraphics[width=0.99\columnwidth]{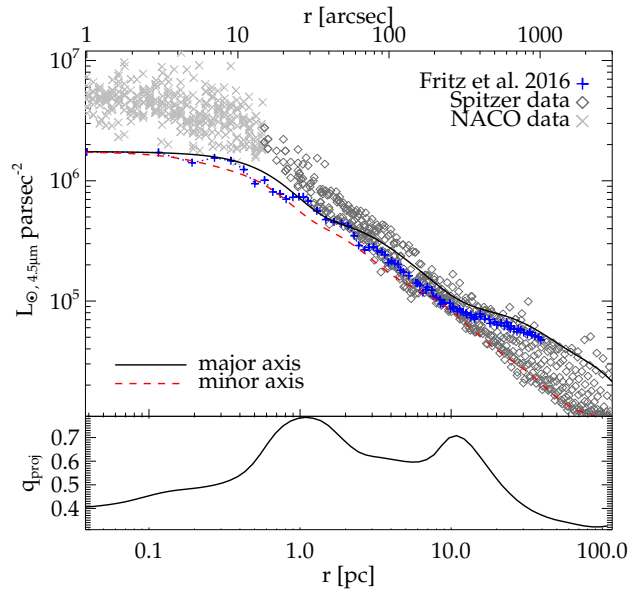}
\caption{Upper Panel: Surface brightness profile derived from a   dust extinction and PAH emission corrected \textit{Spitzer}/IRAC 4.5\,$\mu$m image and NACO $H$-band mosaic for the centre, scaled to the measurements of \citet[blue crosses]{fritz16}. The black full line   denotes the MGE fit along the major axis, and the red dashed line along the  minor axis. Lower panel: Projected axial ratio  $q_\mathrm{proj}$ as a function of $r$.}
	\label{fig:sbh}
	\end{center}
\end{figure}

A comparison with the surface brightness profile of \citet[their Fig. 2]{fritz16} showed that our profile is steeper in the central $\sim$30\,arcsec. A possible reason is the small overlap region of the \textit{Spitzer} and NACO images, and that the \textit{Spitzer} flux could be too high at the centre. Maybe the PAH emission correction of the \textit{Spitzer} image was too low. The mid-infrared dust emission is significant out to almost one arcmin.  
 \cite{fritz16}  used NACO $H$- and $K_S$-band images in  the central $r$\,=\,20\,arcsec. Out to 1\,000\,arcsec\space ($\sim$39\,pc)  they used \textit{Hubble Space Telescope} WFC3 data (M127 and M153 filters) and public VISTA Variables in the Via Lactea Survey  images ($H$ and $K_S$ bands, \citealt{2012A&A...537A.107S}). We lowered the intensities of the central Gaussians by scaling our averaged profile to the one-dimensional flux density profile of \cite{fritz16}. 
 As a result the amplitudes of the inner Gaussians become  smaller, but the outer Gaussians ($\sigma_{\mathrm{MGE}}$\textgreater 100\,arcsec $\sim$4\,pc) are nearly unchanged. We list the components of the Multi-Gaussian Expansion in Table~\ref{tab:sbprofh} and plot the surface brightness profile in Fig.~\ref{fig:sbh} (upper panel). We also show  the   projected axial ratio $q_\mathrm{proj}$ as a function of radius in the lower panel of Fig.~\ref{fig:sbh}. Out to the central 1\,pc, $q_\mathrm{proj}$ is increasing from 0.4 to 0.7.  \cite{sb} found  a mean axial ratio of 0.71$\pm$0.02 for the nuclear star cluster. This is in agreement with our maximum value of $q_\mathrm{proj}$. However,  $q_\mathrm{proj}$ decreases at larger radii, as the contribution from the nuclear stellar disc becomes more important and the light distribution therefore flatter. 
 \begin{table}
\centering
\caption{The Multi-Gaussian-Expansion (MGE) fit parameters for the 4.5\,$\micron$ \textit{Spitzer}/IRAC dust extinction  and PAH  emission corrected image in combination with the  NACO $H$-band mosaic scaled to \textit{Spitzer} flux. $I_\mathrm{scaled} $ is the peak surface brightness used in the dynamical modelling, $\sigma_{\mathrm{MGE}}$ is the standard deviation, and $q_\mathrm{MGE}$ is the axial ratio of the Gaussian components. $I_\mathrm{unscaled} $ is the peak surface brightness before scaling to \citet{fritz16}.}
\label{tab:sbprofh}
\begin{tabular}{cccc}
\hline
$I_\mathrm{scaled} $ &$\sigma_{\mathrm{MGE}}$  &$ q_\mathrm{MGE}$&$I_\mathrm{unscaled} $ \\
$\left[ 10^4\,\mathrm{L}_{\odot,4.5\mu \mathrm{m}}\,\mathrm{pc}^{-2} \right] $  & $\left[ \mathrm{arcsec} \right] $   &&$\left[ 10^4\,\mathrm{L}_{\odot,4.5\mu \mathrm{m}}\,\mathrm{pc}^{-2} \right] $   \\
\hline
$	0.86	$&$	  1.7 		$&$		0.30$	&$		312$\\
$	32.4   	$&$	  10.4  	$&$	   	0.34$&$		164$\\	
$	89.8  	$&$	   15.0   	$&$		0.82$&$		257$\\	
$	18.5	$&$	     52.1  	$&$	  	0.95$	&$		30.0$\\
$	17.0  	$&$	   98  	$&$	  	0.36$&$		29.3$\\	
 $      7.1   	$&$	  154   	$&$	  	0.95$&$		7.4$\\
 $      4.8  	$&$	   637   	$&$	   	0.36$	&$		4.9$\\
$       3.2   	$&$	  2020   	$&$	  	0.30$&$		3.2$\\
$       1.3   	$&$	  4590   	$&$	  	0.81$&$		1.3$\\	
\hline
\end{tabular}
\end{table}

We note that there are three main differences with the surface brightness distribution derived by \cite{isaacanja}: (1) We used  an $H$-band instead of a  $K_S$-band NACO image to avoid ionised  gas emission; (2) We masked  young stars  in the NACO image to match the distribution of stars used as kinematic tracers; and (3) We scaled the central photometry to the flux density  data of \cite{fritz16} to avoid a possible overestimation of the central flux when scaled to the \textit{Spitzer} image. All three changes influence only the central part of the surface brightness distribution, as ionised gas emission and light from young stars are only important in the central parsec. 

The surface brightness profile is deprojected to obtain the three-dimensional spatial stellar light distribution. In general, the deprojection of a surface brightness profile is non-unique \citep[e.g.][]{1987IAUS..127..397R,1988MNRAS.231..285F}, and our deprojection is only one possible solution. The MGE deprojection produces smooth intrinsic densities, which are in agreement with the photometric observations \citep{mgeidl}.

\section{Dynamical Models of the Milky Way nuclear star cluster}
\label{sec:model}
\subsection{Schwarzschild's method}
Orbit-based models or Schwarzschild models are a useful tool to model the dynamics of stellar systems by orbit superposition. 
The first step of Schwarzschild's method is to integrate the equations of motion for  a representative library of stellar orbits in   a gravitational potential $\Phi$. Then the observables for each orbit  are computed, considering projection, PSF convolution and pixel binning. The next step is to find orbital weights to combine the orbits such that they   reproduce the observed data. 
Schwarzschild models are a powerful tool to recover the intrinsic kinematical structure and the underlying gravitational potential \citep{1979ApJ...232..236S,vdv08,remco09}. 
We refer the reader for further details to  \cite{remco08} for implementation and \cite{vdv08} for verification of the triaxial Schwarzschild code. 

\subsubsection{Mass model}
We calculated orbits in the  combined gravitational potential of a supermassive black hole $ \Phi_\bullet $ and the star cluster $\Phi_\star$, inferred from the imaging data. 
As we run triaxial models, there are three intrinsic shape parameters, $q$, $p$, and $u$, for the cluster. The shape parameters characterise the axial ratios for the long, intermediate and short axes $a$, $b$, and $c$. They are defined as $q=c/a$, $p=b/a$, and $u=a'/a$, where $a'$ is the length of the longest axis $a$ projected on the sky. Thus, $u$ represents the compression of $a$ due to  projection on the sky. Each set of axial ratios refers  to a set of viewing angles ($\vartheta, \phi, \psi$, see also \citealt{remco08}).
The surface brightness distribution is deprojected given the intrinsic shape parameters  $q, p, u$, and multiplied with the dynamical mass-to-light ratio \ml\space to get the intrinsic stellar  mass density $\varrho_\star$. 
From Poisson's equation  $\nabla^2 \Phi_\star = 4 \pi G \varrho_\star$ one calculates the gravitational potential. 
We did this for different values of the black hole mass \mbh, dynamical mass-to-light ratio \ml, and different shape parameters. 
 In total our model has five free parameters, \mbh,  \ml, $q$, $p$, and $u$.

Besides the considered stellar population  and the supermassive black hole, there are other components within the nuclear star cluster, which we neglected. We measured a  dynamical mass-to-light ratio, which  combines the stellar mass-to-light ratio with other components. These components are the young stars, ionised gas,  neutral gas, and dark matter.  The young stars are at a distance of  about 0.5\,pc from the supermassive black hole. The lower limit of the total mass of young stars is 12\,000\,$M_\odot$. However, the  total enclosed extended mass in the same region is $\sim$10$^6$\,$M_\odot$ \citep{oh09,isaacanja}, and the mass of the supermassive black hole is 4\,$\times$\,10$^6$\,$M_\odot$. The mass of the young stars is therefore probably negligible. The hot ionised gas has a mass of only a few 100\,$M_\odot$ \citep{ferriere12}, and cannot influence the stellar dynamics significantly. The neutral gas in the circum-nuclear disc may contribute more mass,  estimates range from 10$^4$\,$M_\odot$ \citep{Etxaluze,2012A&A...542L..21R} to  10$^6$\,$M_\odot$ \citep{christopherhcn}, though this is probably the upper limit \citep{genzelreview}. The circum-nuclear disc extends over a distance of about 1\,pc to more than 5\,pc from the centre. At 5\,pc, the total enclosed  mass is  $\sim$10$^7$\,$M_\odot$ \citep{mcginn89,isaacanja}. We decided to neglect the mass distribution of the circum-nuclear disc in our dynamical models, since it is very uncertain, and  makes up only 0.1 to  10 per cent of the enclosed mass. 
The contribution of dark matter to the nuclear star cluster mass is also neglected. \cite{2014IAUS..303..403L} showed that the fraction of dark matter in the central 100\,pc of the Milky Way is about 6.6  per cent, assuming the traditional dark matter profile of \cite{1996ApJ...462..563N}. 

\subsubsection{Orbit library}
The orbit library should be as general as possible and representative for the gravitational potential.  We assumed that the orbits are regular and that   three integrals of motion, $E$, $I_2$, and $I_3$, are conserved. The orbit families consist of box orbits, which can cross the centre and have an average angular momentum of zero, 
and three types of tube orbits, which avoid the centre.
The tube orbits are divided in short-axis-tube orbits, which have non-zero mean angular momentum $\langle L_z \rangle$ around the short axis, outer and inner long-axis-tube orbits, which have non-zero mean angular momentum $\langle L_x \rangle$ around the long axis. 
The orbit grid should sample the entire phase space. It has to  be dense enough to suppress discreteness noise, but integration has to be done in a reasonable amount of computing time.

\begin{figure*}
  \centering  
    \includegraphics[width=0.95\textwidth]{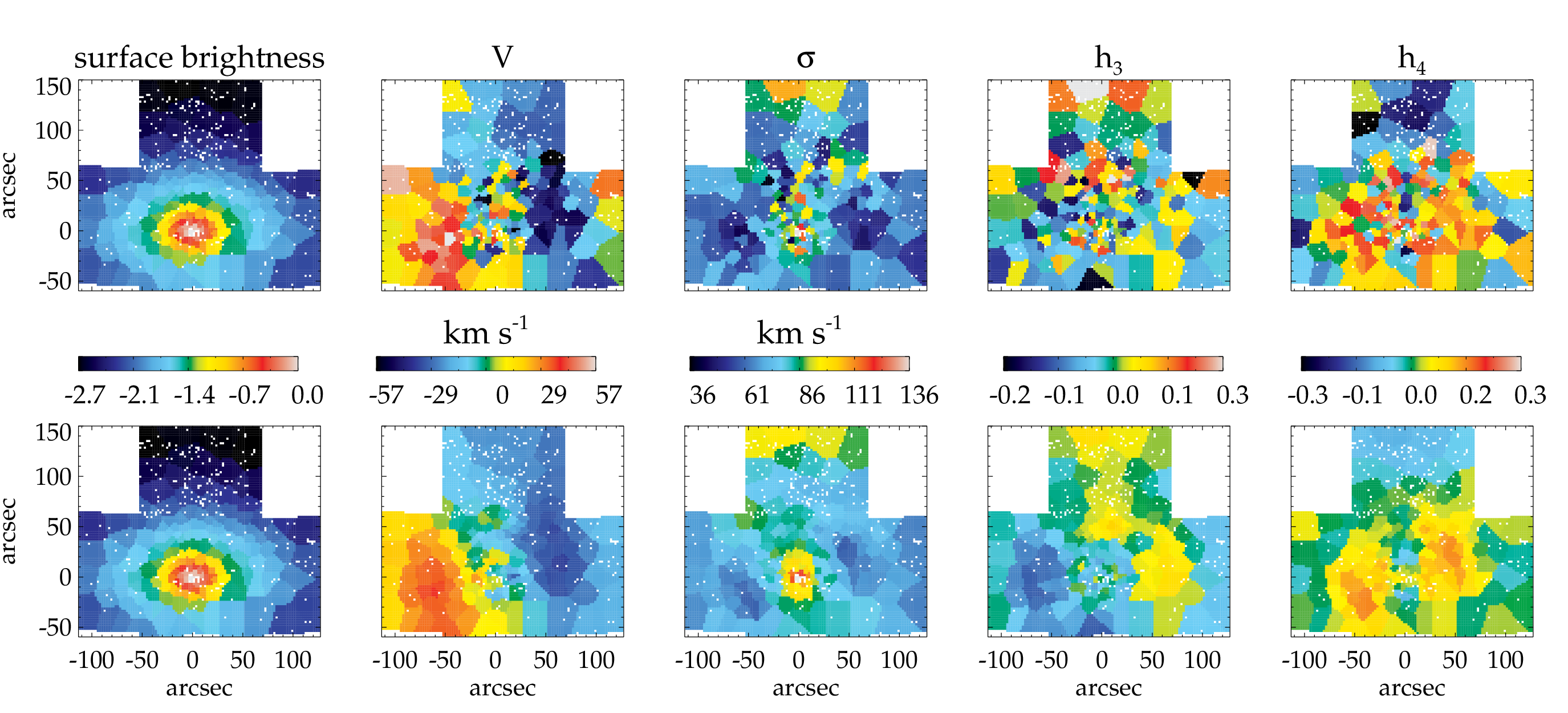}
 \caption{Comparison of the observed stellar surface brightness and kinematics (top row) and the best-fitting Schwarzschild model. The columns denote surface brightness, velocity $V$, velocity dispersion $\sigma$, Gauss-Hermite moments $h_3$, and $h_4$. }
	\label{fig:swbf}
\end{figure*}
\begin{figure}
\centering
     \includegraphics[width=0.99\columnwidth]{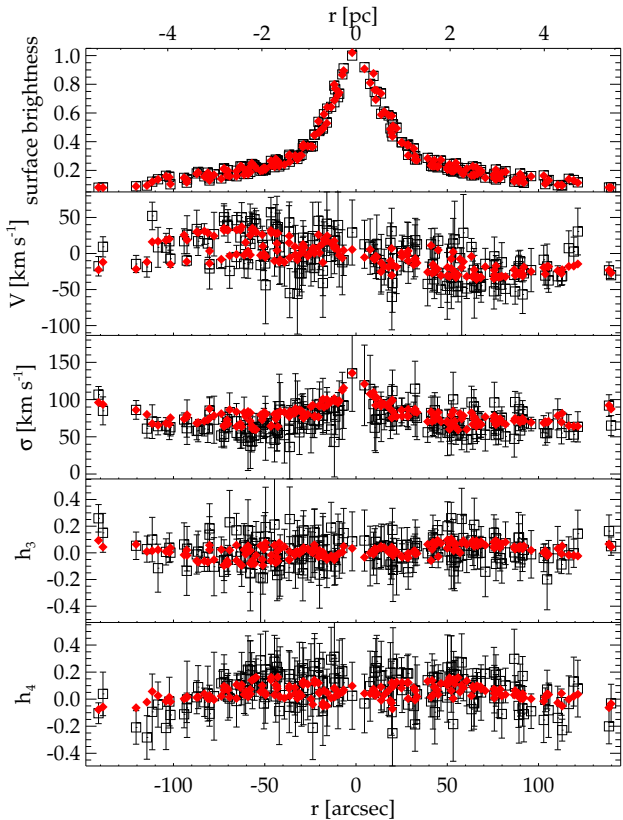}
      \caption{The surface brightness and kinematics of the best-fitting model (red filled diamond symbols) and the data (open black squares and  error bars) as a function of radius. Data points in the Galactic East are plotted at negative radii, Galactic West at positive radii. From top to bottom: surface brightness, velocity $V$, velocity dispersion $\sigma$, and higher Gauss-Hermite moments $h_3$ and $h_4$.  }
	\label{fig:slit}
\end{figure}
We followed \cite{remco08} and  sampled the orbit energy $E$ using a logarithmic grid in radius. Each energy $E$ is linked to the  radius $R_c$ by calculating the potential at $(x, y, z)$ = ($R_c$, 0, 0).  We sampled  $N_E$ = 35 energies calculated from $R_c$ in  logarithmic steps ranging from    $R_c$ = $10^{0.5}$ to $R_c$ = $10^{4.2}$, i.e. 3.16\,arcsec to 4.4\,degree or 0.12\,pc to 616.5\,pc. We note that the outer radius is about 3.5 times the outermost Gaussian $\sigma_{\mathrm{MGE}}$ of the MGE fit. We tested lower values of the inner radius but found consistent results. 
 For each energy, the starting point 
 of an orbit was selected from a linear grid 
 over 14 values each. For details on the orbit sampling we refer to   \cite{remco08}. In total, we have $N_E\,\times\,N_{I_2}\,\times\,N_{I_3} \,=\,  35\,\times\,14\,\times\,14$ = 6860 orbits. Each orbit was integrated  over 200 periods, 
  and sampled on 100\,000 points per orbit. 
 For each orbit we stored  the intrinsic and projected properties. The  projected orbits are stored in a $(x', y', v_z)$ grid, with  PSF convolution and pixel size of the observed data taken into account. The velocities were stored in 183 bins between $-$7.4~$\sigma_{\textrm{max}}$ and +7.4~$\sigma_{\textrm{max}}$. These numbers guarantee a proper sampling of the observed velocity profiles  \citep{1999ApJS..124..383C}. 
 
 \subsubsection{Solving the orbital weight distribution}
The model has to fit the kinematic data,  the intrinsic and the projected mass distribution.
The fit was done by finding a linear combination of the orbits, and solving for orbital weights  $\gamma_i$. Each orbital weight corresponds to a mass on the respective orbit $i$, and the weights $\gamma_i$ are therefore non-negative. We used  the non-negative least-squares (NNLS) algorithm of \cite{1974slsp.book.....L} which was also used by  \cite{1997ApJ...488..702R}, \cite{1998ApJ...493..613V}, and \cite{1999ApJS..124..383C}.
  One of the fitting constraints is to make sure that the model is self-consistent. It is required that  the  orbit superposition   reproduces the intrinsic and projected aperture masses within two per cent, which is the typical accuracy of the observed surface brightness \citep{remco08}. 

\subsection{Constraining the input parameters}

We ran \nmod\space models with different parameter combinations of \mbh, \ml, $q, p, u$. The  black hole mass \mbh\space was sampled in logarithmic steps of   0.2 from 5.5 to 7.5, starting with 6.3  (i.e. \mbh\,$\sim$\,2$\,\times$\,$10^6$\,\msun). The mass-to-light ratio \ml\space was linearly sampled between 0.1 and 2.0 with steps of 0.04, with a starting value of 0.6   (in units of $M_\odot/L_{\odot, 4.5\mu\mathrm{m}}$). The starting model had   ($q, p, u$) = (0.29, 0.84, 0.99). 
We sampled different combinations of ($q, p, u$) with a  step size  of (0.01, 0.02, 0.01), with the boundaries  0.05 \textless $q$ \textless 0.29, 0.40 \textless $p$ \textless 0.99,  and 0.70 \textless $u$ $\leq$ 0.99. 
When  fitting the orbital weights, we already applied the self-consistency criteria  to make sure that the photometry is fitted with a high accuracy of at least two per cent for each model.  The  best fit of the five parameters  \mbh, \ml, $q, p, u$, however, was found    by calculating  the \chisq\space of the kinematic moments. After the starting model we computed different combinations of (\mbh, \ml, $q, p, u$) in the grid. We  used a \chisq\space analysis to find regions with good fits and started new models, with a finer sampling at low \chisq. This was done in several iterations, and it was not necessary to compute each point in the (\mbh, \ml, $q, p, u$) grid.
 
We measured four kinematic moments in 175 spatial  bins of the spectroscopic map in Sect.~\ref{sec:kin}.  In each  bin we measured the velocity $V$, velocity dispersion $\sigma$, and the higher Gauss-Hermite moments  $h_3$, and $h_4$. Thus, the number of observables is the number of spatial bins times the number of kinematic moments, in our case N = 175 $\times$ 4 = 700. We minimised the function  
 
\begin{equation}
\begin{aligned}
\chi^2=\sum_{i=1}^{175}{\left(\frac{V_i-{V_{i}^\text{mod}}}{\epsilon_{V_i}}\right)^2+\left(\frac{\sigma_i-{\sigma_{i}^\text{mod}}}{\epsilon_{\sigma_i}}\right)^2+}\\{\left(\frac{h_{3,i}-{h_{3,i}^\text{mod}}}{\epsilon_{h_{3,i}}}\right)^2+\left(\frac{h_{4,i}-{h_{4,i}^\text{mod}}}{\epsilon_{h_{4,i}}}\right)^2}, 
\label{eq:chi}
\end{aligned}
\end{equation} 
where  $V, \sigma, h_3,$ and $h_4$ denote the kinematic measurements, $\epsilon$  the respective measurement uncertainties,   $V^\mathrm{mod}$, $\sigma^\mathrm{mod}$, $h_3^\mathrm{mod}$, and $h_4^\mathrm{mod}$   the model kinematics, to find the best-fitting model.  

\subsection{Modelling results}
\subsubsection{The best-fitting model}

\begin{figure*}
  \centering  
    \includegraphics[width=0.88\textwidth]{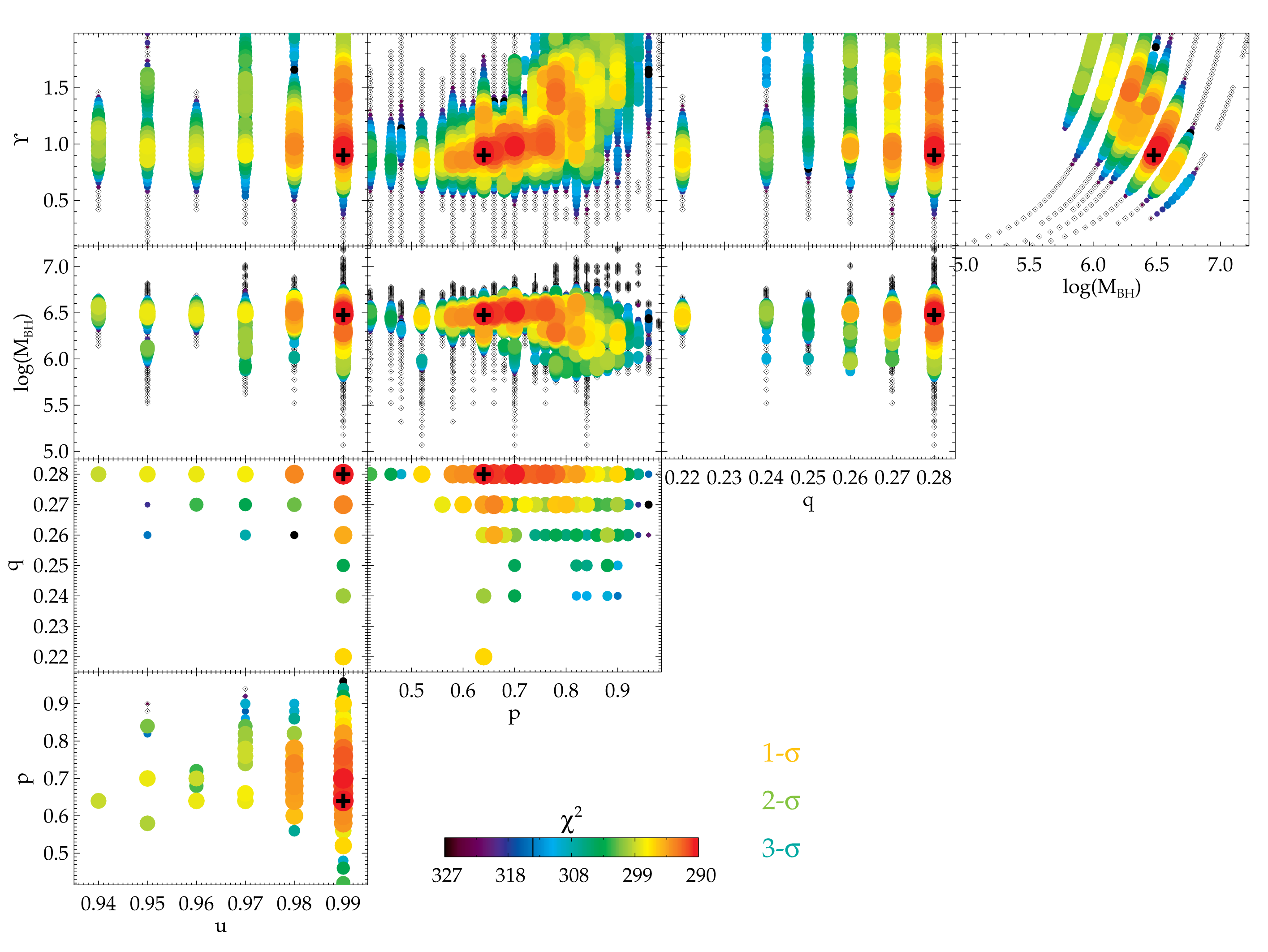}
 \caption{Illustration of the fitted parameter space, \chisq\space was computed with Eq.~\ref{eq:chi}. Each symbol denotes a model, the coloured symbols are models with $\Delta$\chisq\space \textless\space $\sigma_{\chi^2}$ = 37.3, black diamonds are models with $\Delta$\chisq\space\textgreater\space$\sigma_{\chi^2}$. The 1$\sigma$, 2$\sigma$, and 3$\sigma$  colours corresponding to $\Delta \chi^2$ = 5.9, 11.3 and 18.2, are denoted. The black cross denotes the best-fitting model. }
	\label{fig:swchi2}
\end{figure*}
\begin{figure*}
  \centering  
    \includegraphics[width=0.88\textwidth]{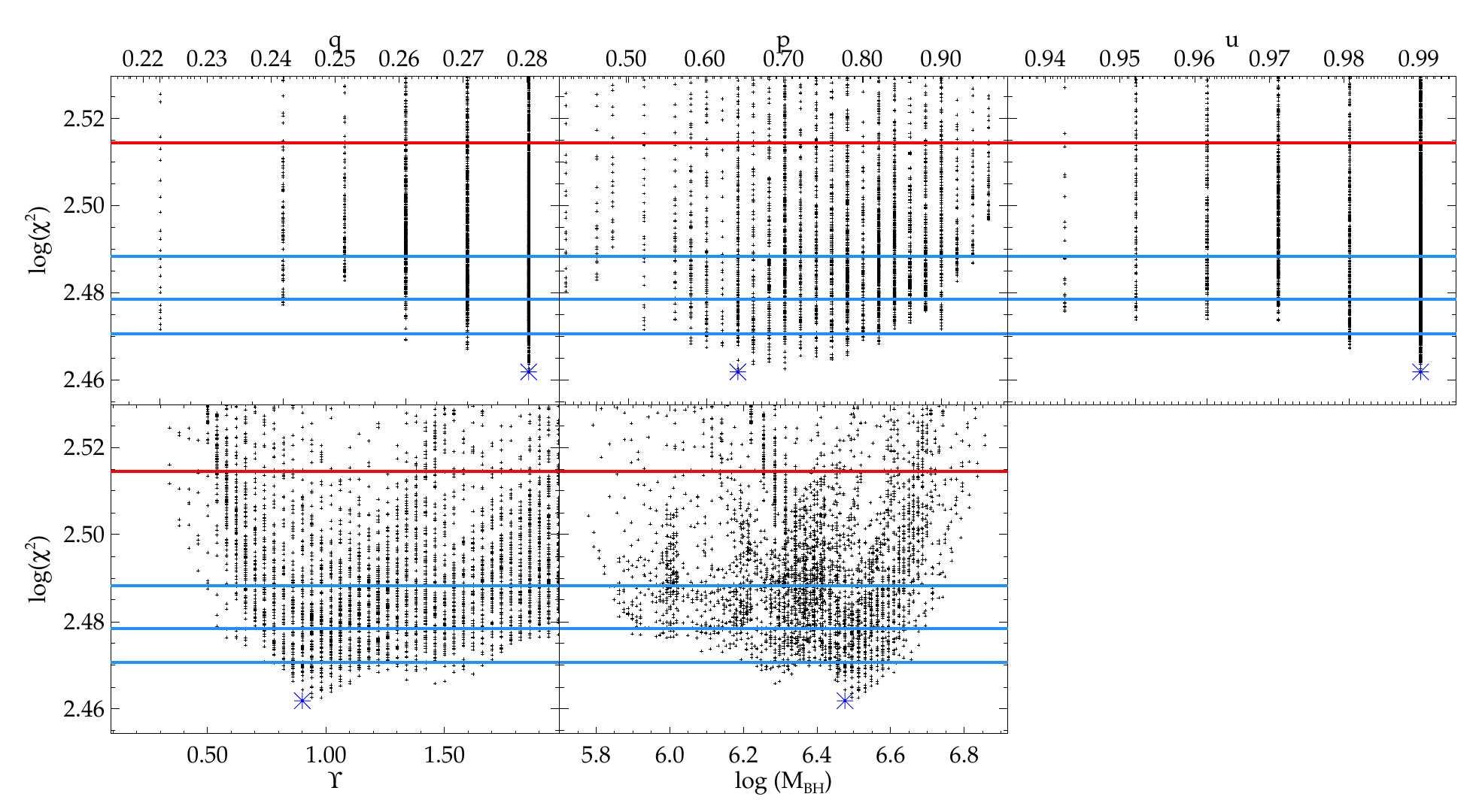}
 \caption{The \chisq\space values plotted against the five free parameters (\mbh, \ml, $q$, $p$, $u$). \chisq\space was computed with Eq.~\ref{eq:chi}. The best-fitting model is denoted as blue asterisk, the 1$\sigma$, 2$\sigma$, and 3$\sigma$ confidence levels, corresponding to $\Delta  \chi^2$ = 5.9, 11.3, and 18.2, are denoted as blue lines. The red line denotes $\sigma_\chi^2$ =  37.3.}	\label{fig:swchi1}
\end{figure*}

Our best-fitting parameters are \mbh\space= \bfbh\,\tentosix\,\msun, \ml\space= \bfml, $q$ = \bfq, $p$ = \bfp, $u$ = \bfu. This corresponds to  best-fitting viewing angles $\vartheta$ = 80\degr,  $\varphi$ = 79\degr, $\psi$ = 91\degr. 
 We show the surface brightness map and the symmetrised kinematic maps in Fig.~\ref{fig:swbf}. The upper row are the data, the lower  row are the maps of the best-fitting model.  The misalignment of the kinematic rotation axis with respect to the photometry, and the perpendicular rotating substructure at $\sim$20\,arcsec\space ($\sim$0.8\,pc) found by \cite{isaacanja} are well reproduced in the model velocity map. We also show the  data and the best-fitting model in Fig.~\ref{fig:slit}, the panels denote surface brightness,  $V$,  $\sigma$, $h_3$ and $h_4$.  
  The surface brightness map is reproduced within  one per cent. The highest and lowest values of $V$, $h_3$, $h_4$, and the lowest values of $\sigma$ in  the data have higher    absolute values than  in the best-fitting model, but are consistent within their uncertainties. 
 The best fit has  \chisq\space = \bfchisq. With   M = 5 fitted parameters and N = 4$\times$175=700 observational constraints, this means \chired\space = \bfchired. That  \chired\space  is less than one is partially due to the large uncertainties of the kinematics, and the fact that the kinematic measurements are correlated. 

We illustrate the distribution of  \chisq\space for the \nmod\space models in Fig.~\ref{fig:swchi2}. We plot each combination of parameters. Red colours denote low \chisq, bluer, smaller symbols denote high \chisq. The black cross denotes the best-fitting model. 
The observed projected flattening $q_\mathrm{MGE}$ of the surface brightness profile  constrains the viewing angles and thus also the intrinsic shape parameters. In particular, a flat $q_\mathrm{MGE}\ll$ 1 means that the stellar system is observed along one of the principal planes \citep{remco08}. 
We obtained as lowest value $q_\mathrm{MGE}$ = 0.30, this limits  the possible projections and puts an upper limit on the parameter $q$. 
Likewise, the value of $u$ = \bfu\space is the boundary value of the grid. 
The values of $q, p,$ and $ u$ denote the respective minimum  values of the radially varying intrinsic axial ratios  $q_\mathrm{intr}$, $p_\mathrm{intr}$ and $u_\mathrm{intr}$ over the entire radial range of the photometry, i.e. the nuclear stellar disc and the embedded nuclear star cluster. 
The upper left panel of Fig.~\ref{fig:swchi2} shows that for each value of $u$, the best-fitting \ml\space is approximately \bfml. A similar behaviour is found with $q$ and $p$. There is only a slight increase of the best-fitting \ml\space with higher values of $p$. At the same time, the best-fitting values of \mbh\space do  not show a strong dependence on   $q, p,$ or $u$ (second row) in the allowed parameter ranges. The intrinsic shape parameters do not influence our best fit for \mbh, as this measurement is mostly made from the inner bins and the outer bins contribute little. The outer bins, however, contribute to  the intrinsic shape fit. 
The supermassive black hole mass and the dynamical mass-to-light ratio are correlated. For higher values of \ml, a lower \mbh\space fits the data.

We show how \chisq\space depends on the different parameters in Fig.~\ref{fig:swchi1}. The best-fitting model, which has the lowest \chisq, is marked as blue asterisk symbol. The blue lines denote the 1$\sigma$, 2$\sigma$, and 3$\sigma$  confidence limits, corresponding to $\Delta \chi^2$ = 5.9, 11.3, and 18.2. The red line illustrates the standard deviation of \chisq\space itself, i.e. $\sqrt{2(\mathrm{N-M})}$ = 37.3, where N = 700, and M = 5. This value was used as confidence limit by \cite{remco09}. In Table \ref{tab:bfres} we  list the 1$\sigma$  and 3$\sigma$ uncertainties.

\begin{table}
\centering
\caption{The best-fitting model results and the 1$\sigma$  and 3$\sigma$  uncertainties, corresponding to $\Delta \chi^2$ = 5.9 and 18.2.  }
\label{tab:bfres}
\begin{tabular}{ccccc}
\hline
parameter& best fit & 1$\sigma$& 3$\sigma$&unit \\
\hline
\noalign{\smallskip}
\mbh  & \bfbh & \bhos& \bhts &\tentosix\space\msun\\
\noalign{\smallskip}
\ml\space  & \bfml & \mlos& \mlts &$M_\odot/L_{\odot, 4.5\mu\mathrm{m}}$\\
\noalign{\smallskip}
$q$ & \bfq & \qos& \qts &...\\
\noalign{\smallskip}
$p$ & \bfp & \pos& \pts &... \\
\noalign{\smallskip}
$u$ & \bfu & \uos& \uts &...\\
\noalign{\smallskip}
\hline
\end{tabular}
\end{table}

\begin{figure}
  \centering  
         \includegraphics[width=0.99\columnwidth]{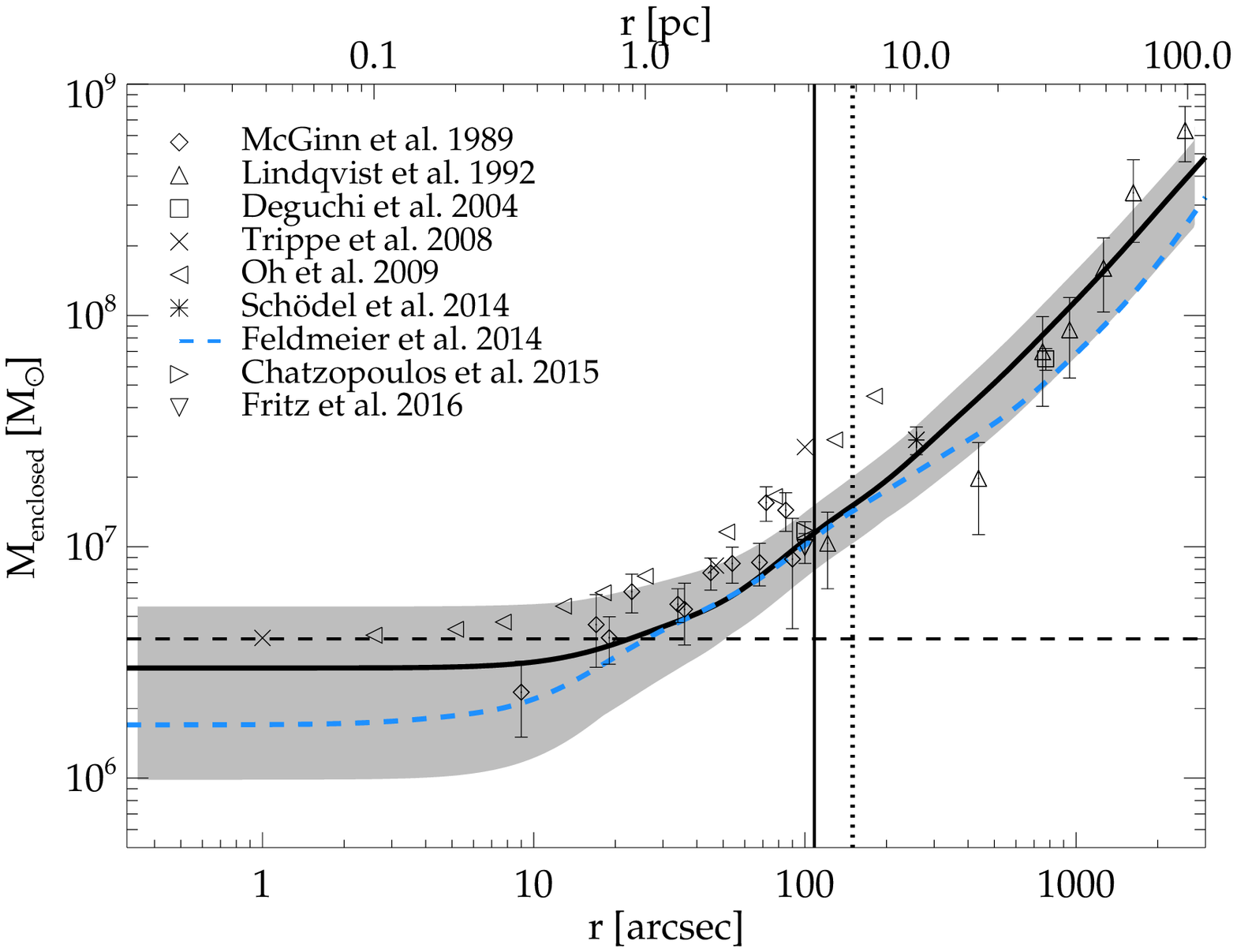}
 \caption{Enclosed total mass in spherical shells  as a function of radius, in units of \msun\space and in logarithmic scaling. The  black line denotes   the enclosed mass with \ml\space= \bfml\space and \mbh\space = \bfbh\,\tentosix\,\msun, the grey shaded contours  the 3$\sigma$ uncertainty. The horizontal line denotes a supermassive black hole with the mass \mbh\space = 4\,\tentosix\,\msun. The vertical, dotted line denotes the outer edge of the kinematic data, the vertical, solid line  the effective radius. We also plot the  results for the enclosed mass from previous studies. We scaled the masses to a distance $R_0$ = 8.0\,kpc,  if the study assumed a different Galactocentric distance: 
\citet[diamonds, assumed $R_0$ = 8.5\,kpc] {mcginn89}, \citet[upward triangles, $R_0$ = 8.5\,kpc]{lindqvist922}, \citet[squares]{deguchi}, \citet[x-symbol]{trippe08}, \citet[leftfacing triangles]{oh09}, \citet[asterisk]{sb}, \citet[blue dashed line]{isaacanja}, \citet[rightfacing triangle, $R_0$ = 8.3\,kpc]{chatzopoulos}, and \citet[downward triangle, $R_0$ = 8.2\,kpc]{fritz16}.}
	\label{fig:encmasss}
\end{figure}
		\begin{figure}
		\centering
	\includegraphics[width=0.95\linewidth]{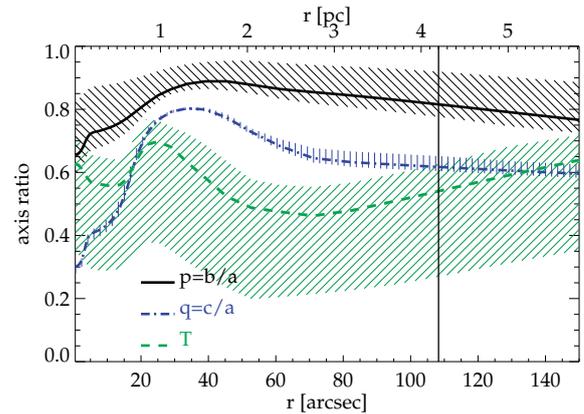} 
	\caption{Intrinsic axial ratios  as a function of radius $r$. The blue dot-dashed line denotes $q=c/a$, the black solid line $p=b/a$,  the green dashed line the triaxiality $T=(1-p^2)/(1-q^2)$, the shaded  regions  the 1$\sigma$ uncertainties, the vertical, solid line   $\reff$ = 4.2\,pc. }
	\label{fig:axisratio}
	\end{figure}
\subsubsection{Mass profile and intrinsic shape }
\label{sec:tridens}
We show the enclosed  total  mass as a function of the projected radius in Fig. \ref{fig:encmasss}, grey shaded contours are the 3$\sigma$ uncertainties. The  mass was computed within spherical shells. We also plot the results of various other studies. Most studies assumed a spherical shape of the cluster, \cite{isaacanja} and \cite{chatzopoulos} assumed axisymmetry. Some of the studies used also different Galactocentric distances, so we scaled the masses using $R_0$ = 8.0\,kpc.  Our results are in agreement with several studies in the central 100\,arcsec. However, we obtain a lower enclosed mass than \cite{trippe08} and \cite{oh09}, who used the Jeans equation for a spherical system to obtain the enclosed mass.
  At larger radii $r$ $\approx$ 400\,arcsec\space ($\sim$15.5\,pc) beyond the reach of our kinematic data, we obtained a  higher mass than \cite{lindqvist922}, but we are in agreement at $r$=750--1600\,arcsec. Their data extend to larger radii, but their assumption of spherical symmetry does no longer hold at such large radii.

 The enclosed total mass within a sphere with the radius $r$ = 8.4\,pc, i.e. about two times the effective radius of the nuclear star cluster, is \mmw\space=  (2.1$\pm$0.7)\,$\times$\,10$^7$\,\msun, here we give the 3$\sigma$ uncertainty.

 The black hole  influences the stellar kinematics only at the centre of the nuclear star cluster.  Out to $r$ = 53\,arcsec\space ($\sim$2\,pc), the best-fitting mass of the black hole (\mbh\space =\,\bfbh\,\tentosix\,\msun) is higher than the enclosed stellar mass of our best-fitting model.   \cite{merritt04} defined the radius of influence of a black hole as the radius where the enclosed stellar mass equals two times the black hole mass. With this definition and a black hole mass of 4\,\tentosix\,\msun, we obtain $\rin$ = (104$^{+56}_{-29}$)\,arcsec, i.e.  approximately (4.0$^{+2.2}_{-1.1}$)\,pc.  
This result is in  agreement with \cite{alexanderrev05}, who found $\rin$ = 3\,pc. The kinematic measurements at larger radii have little influence on the black hole mass measurement, but are important to constrain the orbital structure and dynamical mass-to-light ratio.

The shape of the nuclear star cluster is illustrated in Fig.~\ref{fig:axisratio}. We show the intrinsic axial ratios $q$ and $p$ as a function of radius $r$. The axial ratio $q=c/a$ is low in the centre ($q$=0.3), increases to $q$=0.8 at $r$$\approx$35\,arcsec, and then  decreases to $q$=0.6 at $r$=150\,arcsec.  The  best-fitting shape parameter, $q$=\bfq, is   approximately the central value. The axial ratio $p=b/a$ is also low in the centre ($\sim$0.65), increases to $p$=0.9 at $r$=40\,arcsec, and then decreases to $p$=0.75 at $r$=150\,arcsec. Though, the uncertainty of $p$ is rather high, and the decrease not significant.  Also the  best-fitting axial ratio, $p$=\bfp,  is close to the central value. We  plot the triaxiality $T=(1-p^2)/(1-q^2)$, it varies between 0.45 and 0.7. The triaxiality increases from $r$=70\,arcsec to the outer radius $r$=150\,arcsec.   This may be a signature of the increasing influence of the triaxial Galactic bulge at larger radii  \citep{2017MNRAS.464.3720T}. However, given the large uncertainty of $T$, this increase is not significant. A constant $T$ or a decreasing $T$ are not excluded by the large uncertainties.

\begin{figure}
  \centering  
    \includegraphics[width=0.99\columnwidth]{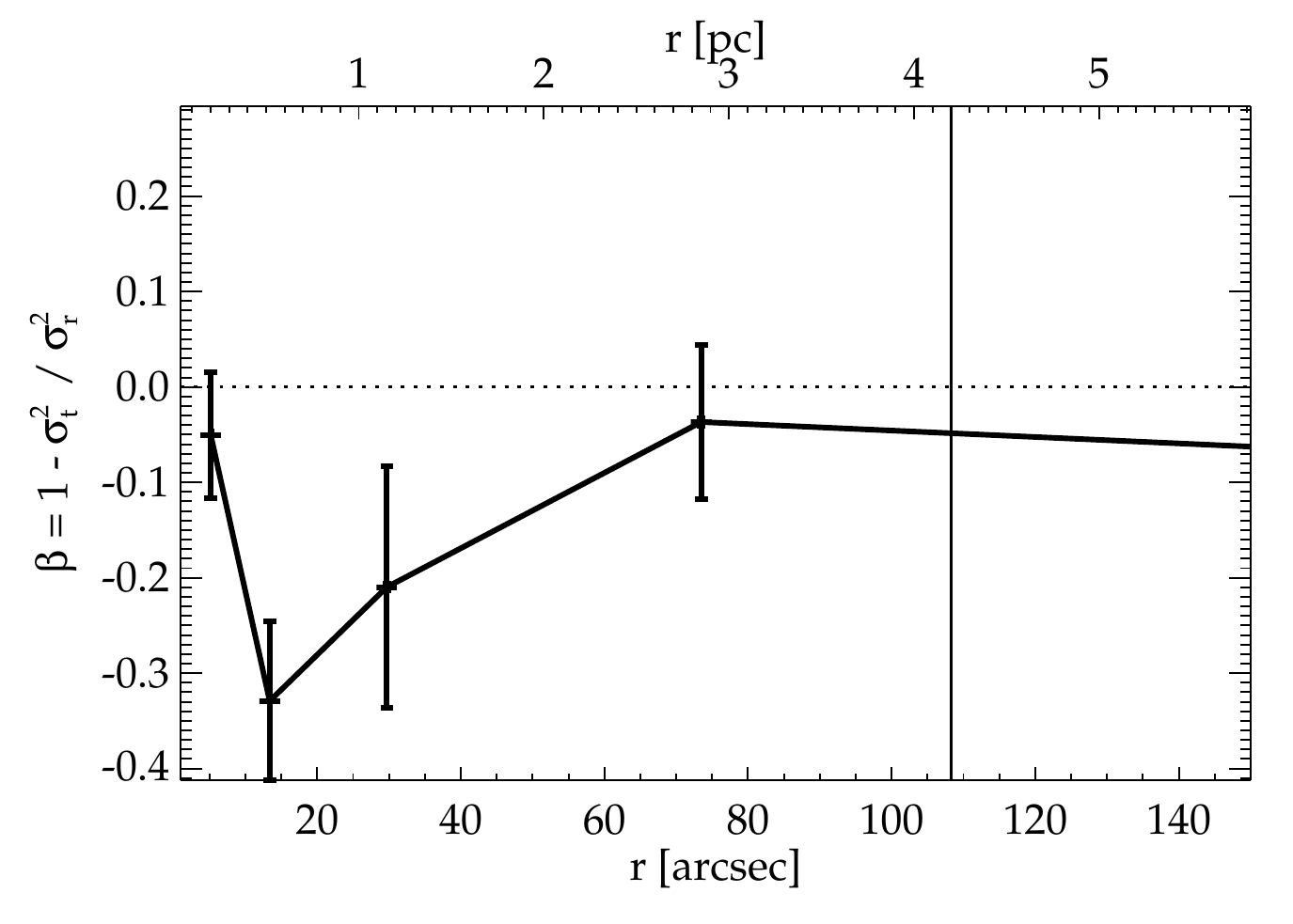}
    \includegraphics[width=0.99\columnwidth]{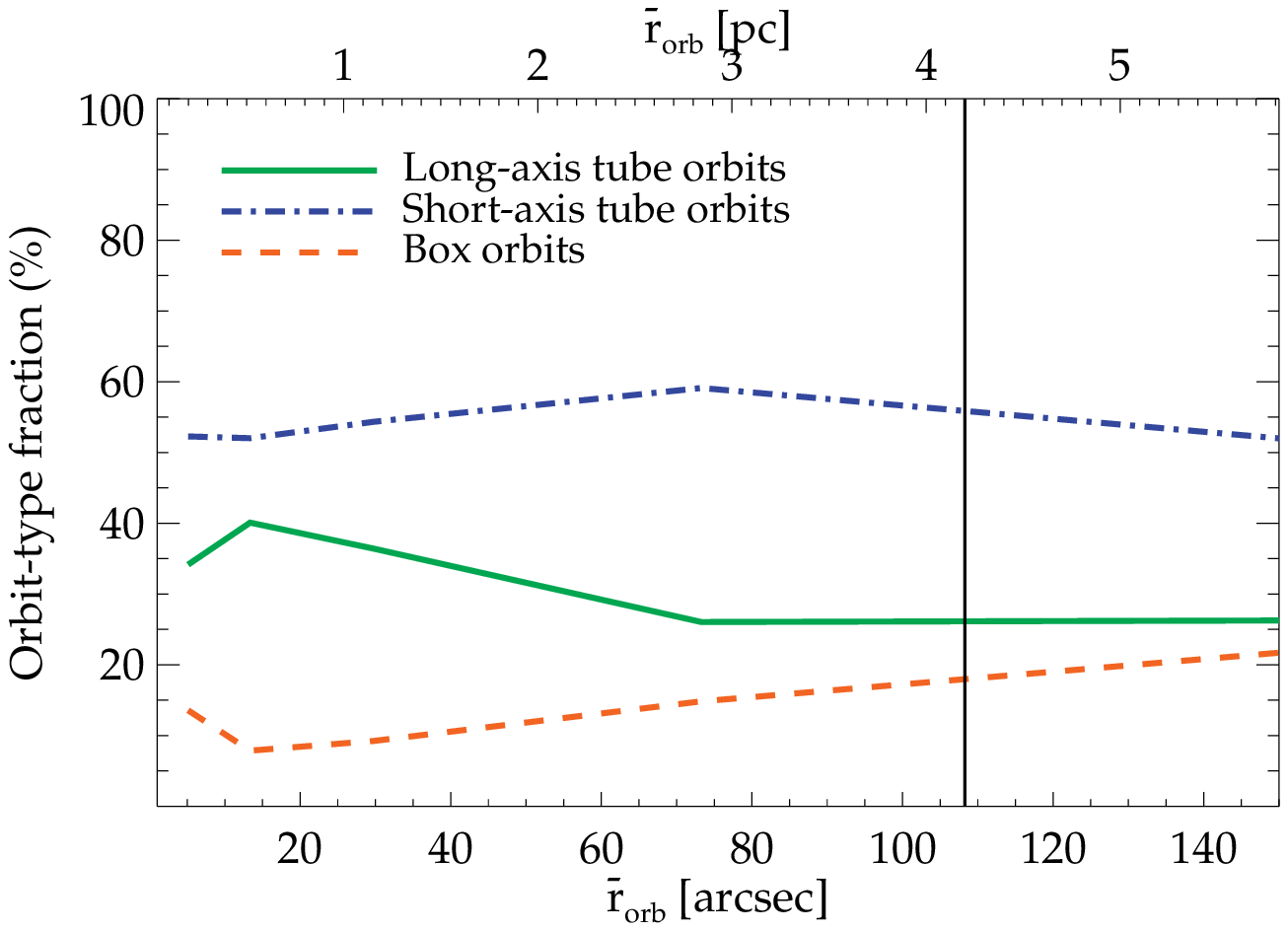}
 \caption{Top: Anisotropy $\beta$ as a function of radius $r$. Negative values denote tangential anisotropy, positive values radial anisotropy.  Bottom:  Fraction of stellar mass per orbit type as a function of average orbit radius $\bar{r}_\text{orb}$. The green, solid line denotes long-axis-tube orbits;  the blue, dot-dashed line  short-axis-tube orbits; the red, dashed line  box orbits; the vertical, solid line  $\reff$ = 4.2\,pc. }
	\label{fig:orbfrac}
\end{figure}

		\begin{figure}
		\centering
	\includegraphics[width=0.95\linewidth]{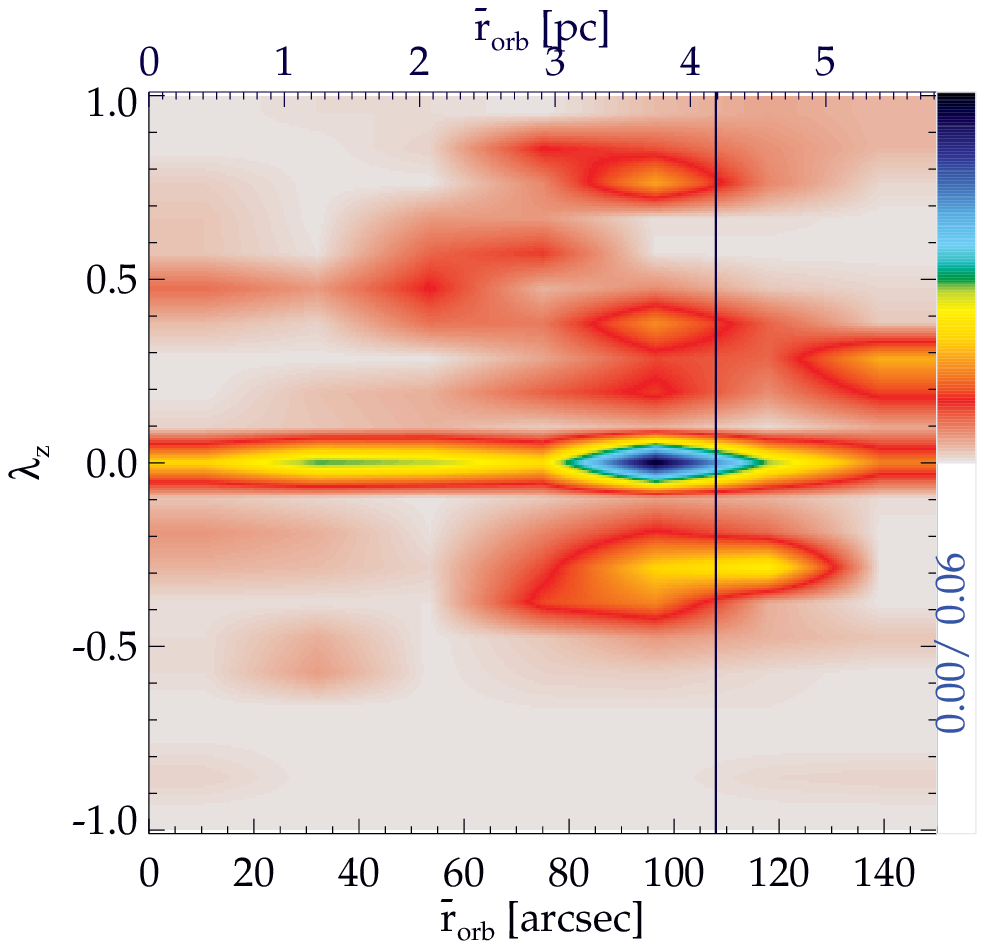} 
	\includegraphics[width=0.95\linewidth]{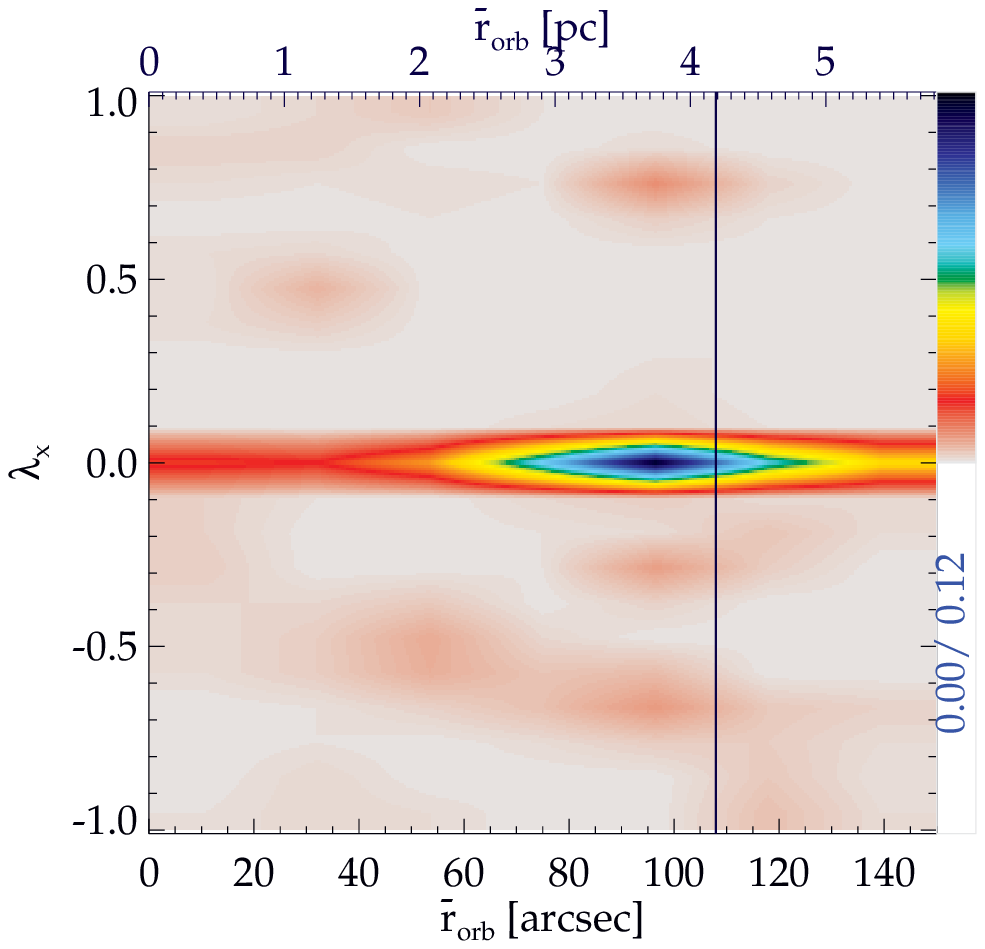} 
	\caption{Orbit density with angular momentum $\lambda_z$ (top), i.e. rotation around the short axis, and $\lambda_x$ (bottom), i.e. rotation around the long axis as a function of average orbit radius $\bar{r}_\text{orb}$. Dark, blue colour indicates higher orbit density.  The vertical line denotes $\reff$ = 4.2\,pc.  }
	\label{fig:lzlx}
	\end{figure}

\subsubsection{Internal dynamics}
The best-fitting model has tangential anisotropy in the centre of the cluster. The value of the anisotropy $\beta$ = $1-\sigma^2_t/\sigma^2_r$\space is negative, where $\sigma_t$ is the tangential velocity dispersion and $\sigma_r$ is the radial velocity dispersion. 
We show the anisotropy $\beta$ as a function of radius in Fig. \ref{fig:orbfrac}, top panel. We plot the mean anisotropy of the models within the 1$\sigma$ uncertainty limit. The uncertainty of $\beta$ is about 0.1. 
 The plot extends to the outer edge of the kinematic data at 150\,arcsec. The cluster kinematics becomes nearly isotropic at  radii $r$ \textgreater 70\,arcsec.

We show the angular momentum distribution of the orbits in Fig. \ref{fig:lzlx}. The colours denote the  density of orbits passing radius $r$ with mean angular momentum $\langle \lambda_z \rangle$ (top panel) or $\langle \lambda_x \rangle$ (bottom panel). The   plot of $\langle \lambda_z \rangle$ denotes rotation about the short z-axis. Orbits with $\langle \lambda_z \rangle \neq$ 0 are contributed by short-axis-tube orbits, while long-axis-tube orbits have $\langle \lambda_z \rangle$ = 0. On the other hand, $\langle \lambda_x \rangle$ denotes rotation about the long x-axis (bottom panel), and orbits with  $\langle \lambda_x \rangle \neq$ 0 are contributed by long-axis-tube orbits. Short-axis-tube orbits have $\langle \lambda_x \rangle$ = 0. Long-axis-tube orbits are most important in the central 20--60\,arcsec\space and at larger radii $r$ $\gtrsim$ 80\,arcsec. Short-axis-tube orbits, which contribute in total more mass than long-axis-tube orbits, are most important at $r$ = 60--140\,arcsec. 
We illustrate the distribution of the stellar mass on the different orbit types also  in Fig. \ref{fig:orbfrac} (bottom panel) as a function of radius.  Most stars (\textgreater 50 per cent) are on short-axis-tube orbits, i.e. they orbit the minor axis.  
Long-axis-tube orbits contribute about 40 per cent to the luminosity and thus stellar mass  in  the central 30\,arcsec. They produce the perpendicular rotating substructure at $r$\,$\approx$\,20\,arcsec\space ($\sim$0.8\,pc) found by \cite{isaacanja}. At larger radii, long-axis-tube orbits contribute only about 30 per cent to the stellar mass. 
Box orbits  contribute little mass in the centre (\textless 10 per cent), but their relative mass increases towards larger radii. At $r$ = 150\,arcsec\space ($\sim$5.8\,pc), they contribute 20 per cent.

\section{Discussion}
\label{sec:discsw}
\subsection{Difference of the resulting black hole mass}

The currently best black hole mass estimate is (4.02 $\pm$ 0.20)\,\tentosix\,\msun, derived  by \cite{2016ApJ...830...17B}  from Keplerian stellar orbits around the supermassive black hole.  Using axisymmetric Jeans models and the same spectroscopic data as our study, \cite{isaacanja} found a  lower value of \mbh\space= (1.7$^{+1.4}_{-1.1}$)\,\tentosix\,\msun. 
Our best fit using triaxial Schwarzschild models is (\bfbh\,\bhos)\,\tentosix\,\msun. This measurement   is consistent with  the direct measurements of \cite{2016ApJ...830...17B} within the 1$\sigma$ uncertainty limit.  The result is also in agreement with the lower black hole mass of  \cite{isaacanja}. 
We derived a 3$\sigma$ lower limit for the black hole of 0.7\,\tentosix\,\msun, and an upper limit of 5.4\,\tentosix\,\msun. 
We briefly discuss  the model degeneracies, possible reasons for the different black hole mass measurements, and why our results are closer to the direct measurement than the black hole mass derived by \cite{isaacanja}. 

\subsubsection{Model degeneracies}
Some model parameters seem to be correlated. This becomes clear when looking at Fig. \ref{fig:swchi2}. The best-fitting value of $p$ apparently increases with increasing dynamical mass-to-light ratio \ml\space (second column of the first row).  However, the value of $p$ has little effect on \mbh, as can be seen in the second column of the second row in  Fig. \ref{fig:swchi2} . With a higher value of $p$, the best-fitting \mbh\space  increases  slightly.  
 At larger $p$, the \chisq-contours of \mbh\space  and \ml\space broaden. This means that  for a more oblate axisymmetric cluster with $p$ closer to one,  \mbh\space and \ml\space are not as well constrained  as with smaller values of $p$. The black hole mass \mbh=4.1\,\tentosix\,\msun\space can be obtained with a higher value of $p$=0.82 and \ml=1.22, this model is within the 1$\sigma$ uncertainties.

The dynamical  mass-to-light ratio \ml\space is inversely correlated with the black hole mass (fourth column of the first row in Fig. \ref{fig:swchi2}). The higher \ml, i.e. the more massive the cluster, the less massive is the black hole. This degeneracy is often obtained in dynamical models.   \cite{2004ApJ...602...66V} found that  the degeneracy of \mbh\space depends on how well the black hole's sphere of influence is resolved. The measurement of \ml\space is better constrained when the data extend to larger radii, provided that  \ml\space is constant over the entire field.  We have several kinematic data bins within the radius of influence of the supermassive black hole, and our data extend to one effective radius. This may not be sufficient to put strong constraints on \ml. 
To get agreement with    the measurement of   (4.02 $\pm$ 0.20)\,\tentosix\,\msun\space   \citep{2016ApJ...830...17B}, 
we would require  a lower value of \ml $\approx$ 0.75, which is within the 1$\sigma$ uncertainties.

\subsubsection{Influence of the surface brightness profile}
The shape of the surface brightness profile is important to estimate the mass of the supermassive black hole. The surface brightness profile has to represent the density of the kinematic tracer. We excluded young stars and ionised gas from the surface brightness profile, as these components contribute little mass compared to the cool, old stars we used as kinematic tracers. 
Excluding these components results in a lower surface brightness and stellar mass in the centre compared to \cite{isaacanja}. 
 The stellar mass we obtain  at $r$ = 48\,arcsec\space ($\sim$1.8\,pc) is 1.3\,\tentosix\,\msun\space less. Our black hole mass is  therefore higher, and closer to the direct measurement of \mbh\space $\approx$ 4\,\tentosix\,\msun. 
 We ran the same axisymmetric models \citep{jam}, using the same kinematic data as \cite{isaacanja}, but our surface brightness distribution from Table \ref{tab:sbprofh}. 
The best fit is obtained with  \mbh\space = ($2.8^{+1.3}_{-0.8}$)\,\tentosix\,\msun, \ml\space = 0.89\,$^{+0.12}_{-0.19}$, and a constant anisotropy of $\beta$ = $-$0.3. This result is in agreement with the triaxial Schwarzschild models, and confirms that the surface brightness profile has a strong influence on the results of the  black hole mass and dynamical mass-to-light ratio.

The deprojection of the surface brightness profile is non-unique and only one possible solution. We applied the MGE method, which  produces smooth intrinsic densities \citep{mgeidl}.  We assumed that the deprojected density is smooth, as this is what we observe in other galaxies, where the clusters are observed from various viewing angles. 

\cite{1997MNRAS.287..543V}  studied how much central density can be hidden in an oblate axisymmetric galaxy without effects on the projected surface brightness. 
They found that the percentage of hidden density in the centre of a Staeckel potential is 0 per cent for inclination $i$=90$^\circ$, and $\lesssim$10 per cent for $i\gtrsim$80$^\circ$ of the total galaxy mass. The percentage of hidden density can  increase in steeper cusps. However, the  effect on the dynamics is still negligible.   
For our models, a hidden density is equivalent to a spatially varying mass-to-light ratio.

\subsubsection{Spatially varying mass-to-light ratio}
We assumed a constant dynamical mass-to-light ratio \ml\space for the Schwarzschild models. 
We obtained  \ml\space = \bfml\space\mlos\space (1$\sigma$  uncertainty). The dynamical mass-to-light ratio combines the stellar mass-to-light ratio with other components, it is sensitive to the presence of  gas or dark matter.

 Our best-fitting value of \ml\space= 0.9 is consistent  with stellar-population studies.  \cite{2014ApJ...797...55N} found \ml\space=0.8--1.2 at  4.6\,$\micron$   for stellar populations  with  ages \textgreater 7.0\,Gyr and metallicities  [$Fe/H$] ranging from -1.0\,dex to +0.3\,dex. For populations with lower metallicity  [$Fe/H$]=--2.18\,dex  and 13\,Gyr age, \ml\space can be even 1.5.  However, for younger stellar populations $\la$5\,Gyr, \cite{2014ApJ...797...55N} obtained lower values \ml\space$\approx$0.6. 
  Our measurement of \ml\space is averaged over the entire  field of the kinematic data.  We cannot exclude that the stellar age or metallicity changes over the range of the kinematic data. Stellar population studies of the red giant population were so far  confined to the central 1--2\,pc \citep[e.g.][]{pfuhl11,dolowfe,kmoslt}. The stars in this region are mostly older than 5\,Gyr and rather metal-rich. Our knowledge of the stellar population at  the outer region of our field is based on only a few bright stars \citep[e.g.][]{blum03,isaacanja}. But these stars are brighter and probably younger than our kinematic tracer stars.   However, the mass-to-light ratio   for old stars in the mid-infrared varies modestly  with age and metallicity in comparison to the optical mass-to-light ratio \citep{meidt}. Therefore we do not expect a change of \ml\space by more than $\sim$0.4 within the cluster.
 Should \ml\space vary with radius, our mass profile (Fig. \ref{fig:encmasss})  would have a different shape. 
For example, if \ml\space was lower in the centre than outside, this would increase \mbh, and there would be less mass in the stellar component. 

However, the stellar mass-to-light ratio may also increase towards the central $r$=0.5\,pc, as massive stellar remnants may migrate  to the centre. 
The mass and distribution of dark stellar remnants, i.e. stellar-mass black holes and neutron stars, in the central parsec of the nuclear star cluster is uncertain. For a top-heavy initial mass function, there could be \textgreater1\,\tentosix\,\msun\space in dark remnants \citep{1993ApJ...408..496M}, though \cite{2010MNRAS.402..519L} found a lower mass of about 1\,$\times$\,10$^5$\,\msun\space for a canonical initial mass function. 
  
  In our models we neglected the mass of  molecular gas in the circum-nuclear disc. The molecular gas may contribute 10$^4 - 10^6$\,\msun.  The gas disc extends from $r$ $\approx$ 1$-$7\,pc along the Galactic plane, but only to $r$ $\approx$ 3\,pc along the minor axis \citep{ferriere12}. Thus, the molecular gas is located in  the  central part of our spectroscopic field, but absent in the North. If the gas contributes significantly to the cluster mass, our assumption of spatially constant \ml\space would be violated, and  \ml\space would be higher than for a stellar component alone.   When we assume the maximum  gas mass of $10^6$\,\msun,  the value of a constant   \ml\space  decreases to about 0.85, which is within our 1$\sigma$ uncertainty limit. 
  
The spatial distribution of dark matter  in the Galactic centre is uncertain. A classical cuspy \cite{1996ApJ...462..563N} dark matter  profile results in a dark matter fraction of  about 6.6  per cent in the central 100\,pc \citep{2014IAUS..303..403L}. However, black hole accretion, dark matter annihilation, and scattering  alter the shape of the dark matter distribution in the Galactic centre.  \cite{2008PhRvD..78h3506V} found that these effects produce a shallower dark matter profile in the central 2\,pc than further out. The  dark matter mass inferred from the classical cusp is reduced by up to 50 per cent in the central 2\,pc.   
The contribution of dark matter to  the nuclear star cluster mass should therefore be negligible. Although the dark matter distribution may be different from the luminous baryonic matter, and the  dynamical  mass-to-light ratio for that reason not spatially constant, the effect on the cluster mass distribution should be only  minor. 

\subsection{Triaxial cluster shape}
Our best-fitting model has axial ratios of $q$\,= $c/a$ = \bfq\,\qos,  $p$ = $b/a$ = \bfp\,\pos, and $u$ = $a'/a$ = \bfu\,\uos.  These axial ratios correspond to viewing angles $\vartheta$ = 80\degr,  $\varphi$ = 79\degr, and $\psi$ = 91\degr. The angle  $\vartheta$ denotes the polar viewing angle, $\varphi$ the azimuthal viewing angle, and  
 $\psi$ is the misalignment angle between photometric major axis and the projected intrinsic long axis \citep{remco08,remco09}.  
For the best-fitting model the angle $\alpha$ between the cluster's major axis  and the line-of-sight    is  about 79\degr.
 The cluster's shape is illustrated in Fig.~\ref{fig:axisratio}, the  triaxiality paramter  $T= (1-p^2)/(1-q^2)$ varies between  0.45 and 0.7, with an average at $T\sim$0.6. An oblate axisymmetric system has $T$ = 0, a prolate axisymmetric system has $T$ = 1. 
 
 Also the Milky Way's bulge is triaxial, the axial ratios are  $q$ = 0.26 and  $p$ = 0.63   \citep{wegg13}. The shape was derived from the density of red clump stars in the central 800\,pc of the bulge. The Milky Way  bulge is much larger than the nuclear star cluster,  and extends out to about 2.5\,kpc.    At the outer edge of our data, at $r$=150\,arcsec, we obtain for the intrinsic shape parameters  $q$=0.60 and $p$=0.75. Both $q$ and $p$ are  higher than found in the bulge. However, they are also decreasing, though the decrease of $p$ is not significant.  
 The bulge has a peanut or X-shape \citep{nataf10,mcwilliam10}. The  angle $\alpha$ between the bulge major axis  and the line-of-sight to the Galactic centre  is  about 27\degr\space  \citep{rattenbury07,wegg13}, while we obtained 79\degr\space  for the nuclear star cluster. There are also indications   for  another bar within the Galactic bulge from star count data  in the inner $|l|\lesssim1\degr\approx$ 140\,pc \citep{2001A&A...379L..44A} or even $|l|\lesssim4\degr\approx$ 560\,pc \citep{2005ApJ...621L.105N}. \cite{2008A&A...489..115R} derived an angle $\alpha\approx$60-75\degr\space for a thick triaxial nuclear bar with axial ratios $q\approx$0.55 and $p\approx$0.75.
   
 One possible scenario for nuclear star cluster formation  is  that  massive star clusters  (10$^5$--10$^7$\,\msun)  formed in the galactic  disc,  migrated to the galaxy's centre and merged \citep{nadine11,2016MNRAS.461.3620G}. 
 Simulations of multiple   star cluster   mergers and of star cluster accretion on a nuclear stellar component can  produce triaxial nuclear star clusters \citep{2004ApJ...610L..13B,hartmann11,perets14}. 
However, so far no  observational study was able to constrain the triaxial shape of nuclear star clusters in general. 
\cite{hartmann11}   constrained the shape of two nuclear star  clusters and  found agreement with an  axisymmetric shape. For the Milky Way nuclear star cluster, $p$ may be as high as 0.94 within the 3$\sigma$ uncertainties. Thus, a nearly axisymmetric shape is consistent with our data.

\subsection{Caveats and Considerations}
\subsubsection{Regime of semi-resolved populations}
We used integrated light spectroscopy to measure the stellar kinematics. This is the common approach for extragalactic systems, which have a distance of several Mpc. The measured kinematics are weighted by the respective luminosities of  different stars. As the Milky Way nuclear star cluster is only 8\,kpc distant, we are in the regime of semi-resolved populations. The brightest stars can be resolved individually, and these stars contribute a large fraction of the flux. In consequence,  individual spatial bins can be dominated by a single star. Instead of measuring the spectrum of an ensemble of stars, one measures a spectrum in which a large percentage of the flux is contributed by one single star. This causes shot noise, and high differences between neighbouring spatial bins. We accounted for this problem by excluding the brightest stars from the spectroscopic map. This method helps to  significantly reduce the intrinsic scatter of the velocity dispersion \citep{nora6388,bianchini15}. We further increased the kinematic uncertainties such  that the data in two neighbouring bins have consistent values within their uncertainties. This helps to prevent that the models fit only stochastic shot noise. Due to the large kinematic uncertainties, the intrinsic shape parameters $q$, $p$, $u$, and the dynamical mass-to-light ratio \ml\space are not very well constrained, and have   large error margins. 

At a distance of only 8\,kpc, also the relative distances of the stars become more important.  A star  located on the near side of the nuclear star cluster, at a distance d = 7.9\,kpc, contributes 1.05 times more flux than a star with the same absolute magnitude at the far side of the cluster, at d = 8.1\,kpc. 
In an extragalactic system, the distance of a star at the near side  and the distance of a star at the far side with respect to the observer are approximately  the same, as the system is farther away. For a galaxy at d = 5\,Mpc,  a relative difference of 200\,pc changes the  flux  only by a factor 1.00008. Even foreground stars that belong to the outer parts of the stellar system contribute roughly the same flux as a star with the same magnitude that is located in the galactic nucleus. 

\subsubsection{Interstellar extinction}
 
 Another observational complication is interstellar extinction in the Galactic centre, which   varies on arcsecond scales \citep{rainer10}. In particular, the field of view of the kinematic data contains the so-called 20-\kms-cloud \citep[M-0.13-0.08, e.g.][]{20kmsref} in the Galactic southwest.  It lies at a projected distance of about 70\,arcsec\space ($\sim$3\,pc)  from the centre, and  probably about 5\,pc in front of Sgr~A* \citep{ferriere12}. This cloud  blocks the light from stars of the nuclear star cluster. We cannot access the kinematics of stars behind this cloud.  There is also interstellar dust  within  a projected distance of 20\,arcsec\space ($\sim$0.8\,pc) from the centre, i.e. within the radius of influence of the black hole.  This dust causes extinction within   the nuclear star cluster by up to 0.8\,mag \citep{chatzopoulos15}. 
As a consequence, the two effects of dimming by distance and by extinction add up and stars that lie on the far side of the nuclear star cluster appear even more faint than the stars on the near side.

\subsubsection{Implications}

Both  the semi-resolved stellar population and the inter-cluster extinction   cause that our observations are biased to the near side of the  nuclear star cluster. As a consequence,   we measured  a lower limit of the velocity dispersion.  
\cite{isaacanja} found that the  velocity dispersion   in the projected radial range 6\,arcsec\space\textless\space$r$ \textless \space20\,arcsec\space is smaller  compared to the velocity dispersion computed from proper motion data of \cite{Rainerpm09}, which is based on resolved stars.  For resolved stars, the velocity dispersion is not weighted by the flux of the stars.  An underestimated velocity dispersion means that the black hole mass measurement is biased to lower values. 

This observational bias also influences the measurements of $V$, $h_3$ and $h_4$. In particular, the cluster may appear compressed along the line-of-sight, and thus the value of $p = b/a$ = \bfp \pos\space may be too low. As a consequence, \ml\space = \bfml \mlos\space would be underestimated (see second column of the first row in Fig. \ref{fig:swchi2}). 
 
 \subsubsection{Influence of   figure rotation}
The Galaxy rotates, and with it the nuclear star cluster. In a non-axisymmetric, rotating system, centrifugal and Coriolis forces play a role.  However, figure rotation and the resulting forces  were not included in our triaxial models. Figure rotation influences the  stellar orbits. The prograde and retrograde tube orbits no longer fill the same volumes, while the box orbits acquire net mean angular momentum  \citep[e.g.][]{heisler82,schwarzschild82,1993RPPh...56..173S,2002MNRAS.333..847S}.  As a result, orbit-based, tumbling, triaxial models are computationally expensive. Other than an early attempt by \cite{1996MNRAS.283..149Z}, no such models have been constructed that take into account kinematic data. It is difficult to predict how our results would change in a rotating model. The inferred orbital structure will be affected (depending on the tumbling speed of the nuclear star cluster), but  our results on the mass distribution are likely to be fairly robust, as the assumption of a constant mass-to-light ratio is  probably more important.

\section{Summary and Outlook}
\label{sec:sumsw}
We constructed for the first time  triaxial orbit-based Schwarzschild models of the Milky Way nuclear star cluster. We used the spectroscopic integrated light maps of \cite{isaacanja} to measure the cluster kinematics of the central 60\,pc$^2$ of the Milky Way. As photometry we used \textit{Spitzer} 4.5$\micron$\space and NACO $H$-band images, and measured a two-dimensional surface brightness distribution. We excluded young stars,  avoided  gas emission and  dark clouds in the photometric data. Our triaxial models were based on the code by \cite{remco08}. 
 Our best-fitting model contains a black hole of mass \mbh\space= (\bfbh\bhos)\,\tentosix\,\msun, a dynamical mass-to-light ratio of \ml\space= (\bfml \mlos)\,\msun/$L_{\odot, 4.5\mu \mathrm{m}}$, and  shape parameters $q$ = \bfq\qos, $p$ = \bfp\pos,  and $u$ = \bfu\uos. 
Our black hole mass measurement is in agreement with the direct measurement of   (4.02 $\pm$ 0.20)\,\tentosix\,\msun\space  by  \cite{2016ApJ...830...17B}.
We obtain a total  cluster mass  \mmw\space=  (2.1$\pm$0.7)\,$\times$\,10$^7$\,\msun within a spherical shell with radius $r$ = 2\,$\times\,\reff$ = 8.4\,pc.  The best-fitting model is tangentially anisotropic in the central $r$ = 0.5-2\,pc of the nuclear star cluster, but  close to isotropic at larger radii.    
The model is able to recover the long-axis rotation in the central $r$ = 0.8\,pc found by \cite{isaacanja}, and the misalignment of the kinematic rotation axis from the photometric minor axis.

There are several possible ways to extend the dynamical models in the future. One way is to include a component for the neutral gas disc inside the nuclear star cluster. If the gas mass is close to the upper limit of 10$^6$\,\msun, the dynamical mass-to-light ratio would probably decrease slightly, and in return would slightly increase the black hole mass. Modelling a spatially varying mass-to-light ratio may provide a better representation of the cluster's intrinsic properties.   Further, proper motions can be included in combination with  discrete line-of-sight velocities, as shown by \cite{ven} and \cite{bosch06} for axisymmetric Schwarzschild models. \cite{watkins13} extended  axisymmetric Jeans models and implemented discrete kinematic data without binning. Using  discrete data means that the stars are not weighted by their luminosities. This prevents the previously discussed bias towards the near side of the cluster.

\section*{Acknowledgements}
GvdV acknowledges partial support from Sonderforschungsbereich SFB 881 ``The Milky Way System'' (subproject A7 and A8) funded by the German Research Foundation. RS acknowledges  funding from the European Research Council under the European Union's Seventh Framework Programme (FP7/2007-2013) / ERC grant agreement n. 614922.
We thank Remco van den Bosch  for  helpful discussions about the project.  We finally thank the anonymous reviewer for useful comments and suggestions.




\bibliographystyle{mn2e_trunc8} 

\footnotesize{
\bibliography{bibs_lt}

\begin{thebibliography}{98}
\expandafter\ifx\csname natexlab\endcsname\relax\def\natexlab#1{#1}\fi

\bibitem[{{Alard}(2001)}]{2001A&A...379L..44A}
{Alard} C., 2001, \aap, 379, L44

\bibitem[{{Alexander}(2005)}]{alexanderrev05}
{Alexander} T., 2005, \physrep, 419, 65

\bibitem[{{Bahcall} \& {Tremaine}(1981)}]{1981ApJ...244..805B}
{Bahcall} J.~N., {Tremaine} S., 1981, \apj, 244, 805

\bibitem[{{Bekki} {et~al.}(2004){Bekki}, {Couch}, {Drinkwater}, \&
  {Shioya}}]{2004ApJ...610L..13B}
{Bekki} K., {Couch} W.~J., {Drinkwater} M.~J., {Shioya} Y., 2004, \apjl, 610,
  L13

\bibitem[{{Bianchini} {et~al.}(2015){Bianchini}, {Norris}, {van de Ven}, \&
  {Schinnerer}}]{bianchini15}
{Bianchini} P., {Norris} M.~A., {van de Ven} G., {Schinnerer} E., 2015, \mnras,
  453, 365

\bibitem[{{Blum} {et~al.}(2003){Blum}, {Ram{\'{\i}}rez}, {Sellgren}, \&
  {Olsen}}]{blum03}
{Blum} R.~D., {Ram{\'{\i}}rez} S.~V., {Sellgren} K., {Olsen} K., 2003, \apj,
  597, 323

\bibitem[{{Boehle} {et~al.}(2016){Boehle}, {Ghez}, {Sch{\"o}del}, {Meyer},
  {Yelda}, {Albers}, {Martinez}, {Becklin}, {Do}, {Lu}, {Matthews}, {Morris},
  {Sitarski}, \& {Witzel}}]{2016ApJ...830...17B}
{Boehle} A., {Ghez} A.~M., {Sch{\"o}del} R., {Meyer} L., {Yelda} S., {Albers}
  S., {Martinez} G.~D., {Becklin} E.~E. {et~al}, 2016, \apj, 830, 17

\bibitem[{{Cappellari}(2002)}]{mgeidl}
{Cappellari} M., 2002, \mnras, 333, 400

\bibitem[{{Cappellari}(2008)}]{jam}
---, 2008, \mnras, 390, 71

\bibitem[{{Cappellari} \& {Emsellem}(2004)}]{ppxf}
{Cappellari} M., {Emsellem} E., 2004, \pasp, 116, 138

\bibitem[{{Chatzopoulos} {et~al.}(2015{\natexlab{a}}){Chatzopoulos}, {Fritz},
  {Gerhard}, {Gillessen}, {Wegg}, {Genzel}, \& {Pfuhl}}]{chatzopoulos}
{Chatzopoulos} S., {Fritz} T.~K., {Gerhard} O., {Gillessen} S., {Wegg} C.,
  {Genzel} R., {Pfuhl} O., 2015{\natexlab{a}}, \mnras, 447, 948

\bibitem[{{Chatzopoulos} {et~al.}(2015{\natexlab{b}}){Chatzopoulos}, {Gerhard},
  {Fritz}, {Wegg}, {Gillessen}, {Pfuhl}, \& {Eisenhauer}}]{chatzopoulos15}
{Chatzopoulos} S., {Gerhard} O., {Fritz} T.~K., {Wegg} C., {Gillessen} S.,
  {Pfuhl} O., {Eisenhauer} F., 2015{\natexlab{b}}, \mnras, 453, 939

\bibitem[{{Christopher} {et~al.}(2005){Christopher}, {Scoville}, {Stolovy}, \&
  {Yun}}]{christopherhcn}
{Christopher} M.~H., {Scoville} N.~Z., {Stolovy} S.~R., {Yun} M.~S., 2005,
  \apj, 622, 346

\bibitem[{{Cretton} {et~al.}(1999){Cretton}, {de Zeeuw}, {van der Marel}, \&
  {Rix}}]{1999ApJS..124..383C}
{Cretton} N., {de Zeeuw} P.~T., {van der Marel} R.~P., {Rix} H.-W., 1999,
  \apjs, 124, 383

\bibitem[{{Deguchi} {et~al.}(2004){Deguchi}, {Imai}, {Fujii}, {Glass}, {Ita},
  {Izumiura}, {Kameya}, {Miyazaki}, {Nakada}, \& {Nakashima}}]{deguchi}
{Deguchi} S., {Imai} H., {Fujii} T., {Glass} I.~S., {Ita} Y., {Izumiura} H.,
  {Kameya} O., {Miyazaki} A. {et~al}, 2004, \pasj, 56, 261

\bibitem[{{Do} {et~al.}(2015){Do}, {Kerzendorf}, {Winsor}, {St{\o}stad},
  {Morris}, {Lu}, \& {Ghez}}]{dolowfe}
{Do} T., {Kerzendorf} W., {Winsor} N., {St{\o}stad} M., {Morris} M.~R., {Lu}
  J.~R., {Ghez} A.~M., 2015, \apj, 809, 143

\bibitem[{{Do} {et~al.}(2013){Do}, {Martinez}, {Yelda}, {Ghez}, {Bullock},
  {Kaplinghat}, {Lu}, {Peter}, \& {Phifer}}]{tuan}
{Do} T., {Martinez} G.~D., {Yelda} S., {Ghez} A., {Bullock} J., {Kaplinghat}
  M., {Lu} J.~R., {Peter} A.~H.~G. {et~al}, 2013, \apjl, 779, L6

\bibitem[{{Drehmer} {et~al.}(2015){Drehmer}, {Storchi-Bergmann}, {Ferrari},
  {Cappellari}, \& {Riffel}}]{2015MNRAS.450..128D}
{Drehmer} D.~A., {Storchi-Bergmann} T., {Ferrari} F., {Cappellari} M., {Riffel}
  R.~A., 2015, \mnras, 450, 128

\bibitem[{{Emsellem} {et~al.}(1994){Emsellem}, {Monnet}, \& {Bacon}}]{mgeeric}
{Emsellem} E., {Monnet} G., {Bacon} R., 1994, \aap, 285, 723

\bibitem[{{Etxaluze} {et~al.}(2011){Etxaluze}, {Smith}, {Tolls}, {Stark}, \&
  {Gonz{\'a}lez-Alfonso}}]{Etxaluze}
{Etxaluze} M., {Smith} H.~A., {Tolls} V., {Stark} A.~A., {Gonz{\'a}lez-Alfonso}
  E., 2011, \aj, 142, 134

\bibitem[{{Feldmeier} {et~al.}(2014){Feldmeier}, {Neumayer}, {Seth},
  {Sch{\"o}del}, {L{\"u}tzgendorf}, {de Zeeuw}, {Kissler-Patig}, {Nishiyama},
  \& {Walcher}}]{isaacanja}
{Feldmeier} A., {Neumayer} N., {Seth} A., {Sch{\"o}del} R., {L{\"u}tzgendorf}
  N., {de Zeeuw} P.~T., {Kissler-Patig} M., {Nishiyama} S. {et~al}, 2014, \aap,
  570, A2

\bibitem[{{Feldmeier-Krause} {et~al.}(2017){Feldmeier-Krause}, {Kerzendorf},
  {Neumayer}, {Sch{\"o}del}, {Nogueras-Lara}, {Do}, {de Zeeuw}, \&
  {Kuntschner}}]{kmoslt}
{Feldmeier-Krause} A., {Kerzendorf} W., {Neumayer} N., {Sch{\"o}del} R.,
  {Nogueras-Lara} F., {Do} T., {de Zeeuw} P.~T., {Kuntschner} H., 2017, \mnras,
  464, 194

\bibitem[{{Feldmeier-Krause} {et~al.}(2015){Feldmeier-Krause}, {Neumayer},
  {Sch{\"o}del}, {Seth}, {Hilker}, {de Zeeuw}, {Kuntschner}, {Walcher},
  {L{\"u}tzgendorf}, \& {Kissler-Patig}}]{kmoset}
{Feldmeier-Krause} A., {Neumayer} N., {Sch{\"o}del} R., {Seth} A., {Hilker} M.,
  {de Zeeuw} P.~T., {Kuntschner} H., {Walcher} C.~J. {et~al}, 2015, \aap, 584,
  A2

\bibitem[{{Ferri{\`e}re}(2012)}]{ferriere12}
{Ferri{\`e}re} K., 2012, \aap, 540, A50

\bibitem[{{Figer} {et~al.}(1999){Figer}, {McLean}, \& {Morris}}]{figer99}
{Figer} D.~F., {McLean} I.~S., {Morris} M., 1999, \apj, 514, 202

\bibitem[{{Franx}(1988)}]{1988MNRAS.231..285F}
{Franx} M., 1988, \mnras, 231, 285

\bibitem[{{Fritz} {et~al.}(2016){Fritz}, {Chatzopoulos}, {Gerhard},
  {Gillessen}, {Genzel}, {Pfuhl}, {Tacchella}, {Eisenhauer}, \&
  {Ott}}]{fritz16}
{Fritz} T.~K., {Chatzopoulos} S., {Gerhard} O., {Gillessen} S., {Genzel} R.,
  {Pfuhl} O., {Tacchella} S., {Eisenhauer} F. {et~al}, 2016, \apj, 821, 44

\bibitem[{{Gao} {et~al.}(2013){Gao}, {Li}, \& {Jiang}}]{2013EP&S...65.1127G}
{Gao} J., {Li} A., {Jiang} B.~W., 2013, Earth, Planets, and Space, 65, 1127

\bibitem[{{Garc{\'{\i}}a-Mar{\'{\i}}n}
  {et~al.}(2011){Garc{\'{\i}}a-Mar{\'{\i}}n}, {Eckart}, {Weiss}, {Witzel},
  {Bremer}, {Zamaninasab}, {Morris}, {Sch{\"o}del}, {Kunneriath}, {Nishiyama},
  {Baganoff}, {Dov{\v c}iak}, {Sabha}, {Duschl}, {Moultaka}, {Karas},
  {Najarro}, {Mu{\v z}i{\'c}}, {Straubmeier}, {Vogel}, {Krips}, \&
  {Wiesemeyer}}]{20kmsref}
{Garc{\'{\i}}a-Mar{\'{\i}}n} M., {Eckart} A., {Weiss} A., {Witzel} G., {Bremer}
  M., {Zamaninasab} M., {Morris} M.~R., {Sch{\"o}del} R. {et~al}, 2011, \apj,
  738, 158

\bibitem[{{Gebhardt} {et~al.}(2000){Gebhardt}, {Richstone}, {Kormendy},
  {Lauer}, {Ajhar}, {Bender}, {Dressler}, {Faber}, {Grillmair}, {Magorrian}, \&
  {Tremaine}}]{2000AJ....119.1157G}
{Gebhardt} K., {Richstone} D., {Kormendy} J., {Lauer} T.~R., {Ajhar} E.~A.,
  {Bender} R., {Dressler} A., {Faber} S.~M. {et~al}, 2000, \aj, 119, 1157

\bibitem[{{Genzel} {et~al.}(2010){Genzel}, {Eisenhauer}, \&
  {Gillessen}}]{genzelreview}
{Genzel} R., {Eisenhauer} F., {Gillessen} S., 2010, Reviews of Modern Physics,
  82, 3121

\bibitem[{{Genzel} {et~al.}(1996){Genzel}, {Thatte}, {Krabbe}, {Kroker}, \&
  {Tacconi-Garman}}]{genzel96}
{Genzel} R., {Thatte} N., {Krabbe} A., {Kroker} H., {Tacconi-Garman} L.~E.,
  1996, \apj, 472, 153

\bibitem[{{Ghez} {et~al.}(2008){Ghez}, {Salim}, {Weinberg}, {Lu}, {Do}, {Dunn},
  {Matthews}, {Morris}, {Yelda}, {Becklin}, {Kremenek}, {Milosavljevic}, \&
  {Naiman}}]{ghez08}
{Ghez} A.~M., {Salim} S., {Weinberg} N.~N., {Lu} J.~R., {Do} T., {Dunn} J.~K.,
  {Matthews} K., {Morris} M.~R. {et~al}, 2008, \apj, 689, 1044

\bibitem[{{Gillessen} {et~al.}(2009){Gillessen}, {Eisenhauer}, {Trippe},
  {Alexander}, {Genzel}, {Martins}, \& {Ott}}]{gillessen09}
{Gillessen} S., {Eisenhauer} F., {Trippe} S., {Alexander} T., {Genzel} R.,
  {Martins} F., {Ott} T., 2009, \apj, 692, 1075

\bibitem[{{Guillard} {et~al.}(2016){Guillard}, {Emsellem}, \&
  {Renaud}}]{2016MNRAS.461.3620G}
{Guillard} N., {Emsellem} E., {Renaud} F., 2016, \mnras, 461, 3620

\bibitem[{{Haller} {et~al.}(1996){Haller}, {Rieke}, {Rieke}, {Tamblyn},
  {Close}, \& {Melia}}]{haller96}
{Haller} J.~W., {Rieke} M.~J., {Rieke} G.~H., {Tamblyn} P., {Close} L., {Melia}
  F., 1996, \apj, 456, 194

\bibitem[{{Hartmann} {et~al.}(2011){Hartmann}, {Debattista}, {Seth},
  {Cappellari}, \& {Quinn}}]{hartmann11}
{Hartmann} M., {Debattista} V.~P., {Seth} A., {Cappellari} M., {Quinn} T.~R.,
  2011, \mnras, 418, 2697

\bibitem[{{Heisler} {et~al.}(1982){Heisler}, {Merritt}, \&
  {Schwarzschild}}]{heisler82}
{Heisler} J., {Merritt} D., {Schwarzschild} M., 1982, \apj, 258, 490

\bibitem[{{Jeans}(1922)}]{jeans}
{Jeans} J.~H., 1922, \mnras, 82, 122

\bibitem[{{Lawson} \& {Hanson}(1974)}]{1974slsp.book.....L}
{Lawson} C.~L., {Hanson} R.~J., 1974, {Solving least squares problems}

\bibitem[{{Linden}(2014)}]{2014IAUS..303..403L}
{Linden} T., 2014, in IAU Symposium, Vol. 303, The Galactic Center: Feeding and
  Feedback in a Normal Galactic Nucleus, {Sjouwerman} L.~O., {Lang} C.~C.,
  {Ott} J., eds., pp. 403--413

\bibitem[{{Lindqvist} {et~al.}(1992){Lindqvist}, {Habing}, \&
  {Winnberg}}]{lindqvist922}
{Lindqvist} M., {Habing} H.~J., {Winnberg} A., 1992, \aap, 259, 118

\bibitem[{{L{\"o}ckmann} {et~al.}(2010){L{\"o}ckmann}, {Baumgardt}, \&
  {Kroupa}}]{2010MNRAS.402..519L}
{L{\"o}ckmann} U., {Baumgardt} H., {Kroupa} P., 2010, \mnras, 402, 519

\bibitem[{{L{\"u}tzgendorf} {et~al.}(2011){L{\"u}tzgendorf}, {Kissler-Patig},
  {Noyola}, {Jalali}, {de Zeeuw}, {Gebhardt}, \& {Baumgardt}}]{nora6388}
{L{\"u}tzgendorf} N., {Kissler-Patig} M., {Noyola} E., {Jalali} B., {de Zeeuw}
  P.~T., {Gebhardt} K., {Baumgardt} H., 2011, \aap, 533, A36

\bibitem[{{Malkin}(2012)}]{distance8kpc}
{Malkin} Z., 2012, ArXiv e-prints

\bibitem[{{McGinn} {et~al.}(1989){McGinn}, {Sellgren}, {Becklin}, \&
  {Hall}}]{mcginn89}
{McGinn} M.~T., {Sellgren} K., {Becklin} E.~E., {Hall} D.~N.~B., 1989, \apj,
  338, 824

\bibitem[{{McWilliam} \& {Zoccali}(2010)}]{mcwilliam10}
{McWilliam} A., {Zoccali} M., 2010, \apj, 724, 1491

\bibitem[{{Meidt} {et~al.}(2014){Meidt}, {Schinnerer}, {van de Ven},
  {Zaritsky}, {Peletier}, {Knapen}, {Sheth}, {Regan}, {Querejeta},
  {Mu{\~n}oz-Mateos}, {Kim}, {Hinz}, {Gil de Paz}, {Athanassoula}, {Bosma},
  {Buta}, {Cisternas}, {Ho}, {Holwerda}, {Skibba}, {Laurikainen}, {Salo},
  {Gadotti}, {Laine}, {Erroz-Ferrer}, {Comer{\'o}n}, {Men{\'e}ndez-Delmestre},
  {Seibert}, \& {Mizusawa}}]{meidt}
{Meidt} S.~E., {Schinnerer} E., {van de Ven} G., {Zaritsky} D., {Peletier} R.,
  {Knapen} J.~H., {Sheth} K., {Regan} M. {et~al}, 2014, \apj, 788, 144

\bibitem[{{Merritt}(2004)}]{merritt04}
{Merritt} D., 2004, Coevolution of Black Holes and Galaxies, 263

\bibitem[{{Morris}(1993)}]{1993ApJ...408..496M}
{Morris} M., 1993, \apj, 408, 496

\bibitem[{{Nataf} {et~al.}(2010){Nataf}, {Udalski}, {Gould}, {Fouqu{\'e}}, \&
  {Stanek}}]{nataf10}
{Nataf} D.~M., {Udalski} A., {Gould} A., {Fouqu{\'e}} P., {Stanek} K.~Z., 2010,
  \apjl, 721, L28

\bibitem[{{Navarro} {et~al.}(1996){Navarro}, {Frenk}, \&
  {White}}]{1996ApJ...462..563N}
{Navarro} J.~F., {Frenk} C.~S., {White} S.~D.~M., 1996, \apj, 462, 563

\bibitem[{{Neumayer} {et~al.}(2011){Neumayer}, {Walcher}, {Andersen},
  {S{\'a}nchez}, {B{\"o}ker}, \& {Rix}}]{nadine11}
{Neumayer} N., {Walcher} C.~J., {Andersen} D., {S{\'a}nchez} S.~F., {B{\"o}ker}
  T., {Rix} H.-W., 2011, \mnras, 413, 1875

\bibitem[{{Nishiyama} {et~al.}(2005){Nishiyama}, {Nagata}, {Baba}, {Haba},
  {Kadowaki}, {Kato}, {Kurita}, {Nagashima}, {Nagayama}, {Murai}, {Nakajima},
  {Tamura}, {Nakaya}, {Sugitani}, {Naoi}, {Matsunaga}, {Tanab{\'e}},
  {Kusakabe}, \& {Sato}}]{2005ApJ...621L.105N}
{Nishiyama} S., {Nagata} T., {Baba} D., {Haba} Y., {Kadowaki} R., {Kato} D.,
  {Kurita} M., {Nagashima} C. {et~al}, 2005, \apjl, 621, L105

\bibitem[{{Norris} {et~al.}(2014){Norris}, {Meidt}, {Van de Ven}, {Schinnerer},
  {Groves}, \& {Querejeta}}]{2014ApJ...797...55N}
{Norris} M.~A., {Meidt} S., {Van de Ven} G., {Schinnerer} E., {Groves} B.,
  {Querejeta} M., 2014, \apj, 797, 55

\bibitem[{{Oh} {et~al.}(2009){Oh}, {Kim}, \& {Figer}}]{oh09}
{Oh} S., {Kim} S.~S., {Figer} D.~F., 2009, Journal of Korean Astronomical
  Society, 42, 17

\bibitem[{{Oort}(1977)}]{1977ARA&A..15..295O}
{Oort} J.~H., 1977, \araa, 15, 295

\bibitem[{{Paumard} {et~al.}(2006){Paumard}, {Genzel}, {Martins}, {Nayakshin},
  {Beloborodov}, {Levin}, {Trippe}, {Eisenhauer}, {Ott}, {Gillessen}, {Abuter},
  {Cuadra}, {Alexander}, \& {Sternberg}}]{paumard06}
{Paumard} T., {Genzel} R., {Martins} F., {Nayakshin} S., {Beloborodov} A.~M.,
  {Levin} Y., {Trippe} S., {Eisenhauer} F. {et~al}, 2006, \apj, 643, 1011

\bibitem[{{Paumard} {et~al.}(2004){Paumard}, {Maillard}, \&
  {Morris}}]{paumard04}
{Paumard} T., {Maillard} J.-P., {Morris} M., 2004, \aap, 426, 81

\bibitem[{{Perets} \& {Mastrobuono-Battisti}(2014)}]{perets14}
{Perets} H.~B., {Mastrobuono-Battisti} A., 2014, \apjl, 784, L44

\bibitem[{{Pfuhl} {et~al.}(2011){Pfuhl}, {Fritz}, {Zilka}, {Maness},
  {Eisenhauer}, {Genzel}, {Gillessen}, {Ott}, {Dodds-Eden}, \&
  {Sternberg}}]{pfuhl11}
{Pfuhl} O., {Fritz} T.~K., {Zilka} M., {Maness} H., {Eisenhauer} F., {Genzel}
  R., {Gillessen} S., {Ott} T. {et~al}, 2011, \apj, 741, 108

\bibitem[{{Rattenbury} {et~al.}(2007){Rattenbury}, {Mao}, {Sumi}, \&
  {Smith}}]{rattenbury07}
{Rattenbury} N.~J., {Mao} S., {Sumi} T., {Smith} M.~C., 2007, \mnras, 378, 1064

\bibitem[{{Reid} \& {Brunthaler}(2004)}]{2004reid}
{Reid} M.~J., {Brunthaler} A., 2004, \apj, 616, 872

\bibitem[{{Requena-Torres} {et~al.}(2012){Requena-Torres}, {G{\"u}sten},
  {Wei{\ss}}, {Harris}, {Mart{\'{\i}}n-Pintado}, {Stutzki}, {Klein},
  {Heyminck}, \& {Risacher}}]{2012A&A...542L..21R}
{Requena-Torres} M.~A., {G{\"u}sten} R., {Wei{\ss}} A., {Harris} A.~I.,
  {Mart{\'{\i}}n-Pintado} J., {Stutzki} J., {Klein} B., {Heyminck} S. {et~al},
  2012, \aap, 542, L21

\bibitem[{{Rieke} \& {Rieke}(1988)}]{1988ApJ...330L..33R}
{Rieke} G.~H., {Rieke} M.~J., 1988, \apjl, 330, L33

\bibitem[{{Rix} {et~al.}(1997){Rix}, {de Zeeuw}, {Cretton}, {van der Marel}, \&
  {Carollo}}]{1997ApJ...488..702R}
{Rix} H.-W., {de Zeeuw} P.~T., {Cretton} N., {van der Marel} R.~P., {Carollo}
  C.~M., 1997, \apj, 488, 702

\bibitem[{{Rodriguez-Fernandez} \& {Combes}(2008)}]{2008A&A...489..115R}
{Rodriguez-Fernandez} N.~J., {Combes} F., 2008, \aap, 489, 115

\bibitem[{{Rybicki}(1987)}]{1987IAUS..127..397R}
{Rybicki} G.~B., 1987, in IAU Symposium, Vol. 127, Structure and Dynamics of
  Elliptical Galaxies, {de Zeeuw} P.~T., ed., p. 397

\bibitem[{{Saito} {et~al.}(2012){Saito}, {Hempel}, {Minniti}, {Lucas},
  {Rejkuba}, {Toledo}, {Gonzalez}, {Alonso-Garc{\'{\i}}a}, {Irwin},
  {Gonzalez-Solares}, {Hodgkin}, {Lewis}, {Cross}, {Ivanov}, {Kerins},
  {Emerson}, {Soto}, {Am{\^o}res}, {Gurovich}, {D{\'e}k{\'a}ny}, {Angeloni},
  {Beamin}, {Catelan}, {Padilla}, {Zoccali}, {Pietrukowicz}, {Moni Bidin},
  {Mauro}, {Geisler}, {Folkes}, {Sale}, {Borissova}, {Kurtev}, {Ahumada},
  {Alonso}, {Adamson}, {Arias}, {Bandyopadhyay}, {Barb{\'a}}, {Barbuy},
  {Baume}, {Bedin}, {Bellini}, {Benjamin}, {Bica}, {Bonatto}, {Bronfman},
  {Carraro}, {Chen{\`e}}, {Clari{\'a}}, {Clarke}, {Contreras}, {Corvill{\'o}n},
  {de Grijs}, {Dias}, {Drew}, {Fari{\~n}a}, {Feinstein},
  {Fern{\'a}ndez-Laj{\'u}s}, {Gamen}, {Gieren}, {Goldman},
  {Gonz{\'a}lez-Fern{\'a}ndez}, {Grand}, {Gunthardt}, {Hambly}, {Hanson},
  {He{\l}miniak}, {Hoare}, {Huckvale}, {Jord{\'a}n}, {Kinemuchi}, {Longmore},
  {L{\'o}pez-Corredoira}, {Maccarone}, {Majaess}, {Mart{\'{\i}}n}, {Masetti},
  {Mennickent}, {Mirabel}, {Monaco}, {Morelli}, {Motta}, {Palma}, {Parisi},
  {Parker}, {Pe{\~n}aloza}, {Pietrzy{\'n}ski}, {Pignata}, {Popescu}, {Read},
  {Rojas}, {Roman-Lopes}, {Ruiz}, {Saviane}, {Schreiber}, {Schr{\"o}der},
  {Sharma}, {Smith}, {Sodr{\'e}}, {Stead}, {Stephens}, {Tamura}, {Tappert},
  {Thompson}, {Valenti}, {Vanzi}, {Walton}, {Weidmann}, \&
  {Zijlstra}}]{2012A&A...537A.107S}
{Saito} R.~K., {Hempel} M., {Minniti} D., {Lucas} P.~W., {Rejkuba} M., {Toledo}
  I., {Gonzalez} O.~A., {Alonso-Garc{\'{\i}}a} J. {et~al}, 2012, \aap, 537,
  A107

\bibitem[{{Sch{\"o}del} {et~al.}(2014){Sch{\"o}del}, {Feldmeier}, {Kunneriath},
  {Stolovy}, {Neumayer}, {Amaro-Seoane}, \& {Nishiyama}}]{sb}
{Sch{\"o}del} R., {Feldmeier} A., {Kunneriath} D., {Stolovy} S., {Neumayer} N.,
  {Amaro-Seoane} P., {Nishiyama} S., 2014, \aap, 566, A47

\bibitem[{{Sch{\"o}del} {et~al.}(2009){Sch{\"o}del}, {Merritt}, \&
  {Eckart}}]{Rainerpm09}
{Sch{\"o}del} R., {Merritt} D., {Eckart} A., 2009, \aap, 502, 91

\bibitem[{{Sch{\"o}del} {et~al.}(2010){Sch{\"o}del}, {Najarro}, {Muzic}, \&
  {Eckart}}]{rainer10}
{Sch{\"o}del} R., {Najarro} F., {Muzic} K., {Eckart} A., 2010, \aap, 511, A18

\bibitem[{{Schwarz\-schild}(1979)}]{1979ApJ...232..236S}
{Schwarz\-schild} M., 1979, \apj, 232, 236

\bibitem[{{Schwarzschild}(1982)}]{schwarzschild82}
{Schwarzschild} M., 1982, \apj, 263, 599

\bibitem[{{Scoville} {et~al.}(2003){Scoville}, {Stolovy}, {Rieke},
  {Christopher}, \& {Yusef-Zadeh}}]{scoville03}
{Scoville} N.~Z., {Stolovy} S.~R., {Rieke} M., {Christopher} M., {Yusef-Zadeh}
  F., 2003, \apj, 594, 294

\bibitem[{{Sellgren} {et~al.}(1990){Sellgren}, {McGinn}, {Becklin}, \&
  {Hall}}]{sellgren90}
{Sellgren} K., {McGinn} M.~T., {Becklin} E.~E., {Hall} D.~N., 1990, \apj, 359,
  112

\bibitem[{{Sellwood} \& {Wilkinson}(1993)}]{1993RPPh...56..173S}
{Sellwood} J.~A., {Wilkinson} A., 1993, Reports on Progress in Physics, 56, 173

\bibitem[{{Siopis} {et~al.}(2009){Siopis}, {Gebhardt}, {Lauer}, {Kormendy},
  {Pinkney}, {Richstone}, {Faber}, {Tremaine}, {Aller}, {Bender}, {Bower},
  {Dressler}, {Filippenko}, {Green}, {Ho}, \&
  {Magorrian}}]{2009ApJ...693..946S}
{Siopis} C., {Gebhardt} K., {Lauer} T.~R., {Kormendy} J., {Pinkney} J.,
  {Richstone} D., {Faber} S.~M., {Tremaine} S. {et~al}, 2009, \apj, 693, 946

\bibitem[{{Skokos} {et~al.}(2002){Skokos}, {Patsis}, \&
  {Athanassoula}}]{2002MNRAS.333..847S}
{Skokos} C., {Patsis} P.~A., {Athanassoula} E., 2002, \mnras, 333, 847

\bibitem[{{Stolovy} {et~al.}(2006){Stolovy}, {Ramirez}, {Arendt}, {Cotera},
  {Yusef-Zadeh}, {Law}, {Gezari}, {Sellgren}, {Karr}, {Moseley}, \&
  {Smith}}]{iracim}
{Stolovy} S., {Ramirez} S., {Arendt} R.~G., {Cotera} A., {Yusef-Zadeh} F.,
  {Law} C., {Gezari} D., {Sellgren} K. {et~al}, 2006, Journal of Physics
  Conference Series, 54, 176

\bibitem[{{Trippe} {et~al.}(2008){Trippe}, {Gillessen}, {Gerhard}, {Bartko},
  {Fritz}, {Maness}, {Eisenhauer}, {Martins}, {Ott}, {Dodds-Eden}, \&
  {Genzel}}]{trippe08}
{Trippe} S., {Gillessen} S., {Gerhard} O.~E., {Bartko} H., {Fritz} T.~K.,
  {Maness} H.~L., {Eisenhauer} F., {Martins} F. {et~al}, 2008, \aap, 492, 419

\bibitem[{{Tsatsi} {et~al.}(2017){Tsatsi}, {Mastrobuono-Battisti}, {van de
  Ven}, {Perets}, {Bianchini}, \& {Neumayer}}]{2017MNRAS.464.3720T}
{Tsatsi} A., {Mastrobuono-Battisti} A., {van de Ven} G., {Perets} H.~B.,
  {Bianchini} P., {Neumayer} N., 2017, \mnras, 464, 3720

\bibitem[{{Valluri} {et~al.}(2005){Valluri}, {Ferrarese}, {Merritt}, \&
  {Joseph}}]{2005ApJ...628..137V}
{Valluri} M., {Ferrarese} L., {Merritt} D., {Joseph} C.~L., 2005, \apj, 628,
  137

\bibitem[{{Valluri} {et~al.}(2004){Valluri}, {Merritt}, \&
  {Emsellem}}]{2004ApJ...602...66V}
{Valluri} M., {Merritt} D., {Emsellem} E., 2004, \apj, 602, 66

\bibitem[{{van de Ven} {et~al.}(2008){van de Ven}, {de Zeeuw}, \& {van den
  Bosch}}]{vdv08}
{van de Ven} G., {de Zeeuw} P.~T., {van den Bosch} R.~C.~E., 2008, \mnras, 385,
  614

\bibitem[{{van de Ven} {et~al.}(2006){van de Ven}, {van den Bosch}, {Verolme},
  \& {de Zeeuw}}]{ven}
{van de Ven} G., {van den Bosch} R.~C.~E., {Verolme} E.~K., {de Zeeuw} P.~T.,
  2006, \aap, 445, 513

\bibitem[{{van den Bosch}(1997)}]{1997MNRAS.287..543V}
{van den Bosch} F.~C., 1997, \mnras, 287, 543

\bibitem[{{van den Bosch} {et~al.}(2006){van den Bosch}, {de Zeeuw},
  {Gebhardt}, {Noyola}, \& {van de Ven}}]{bosch06}
{van den Bosch} R., {de Zeeuw} T., {Gebhardt} K., {Noyola} E., {van de Ven} G.,
  2006, \apj, 641, 852

\bibitem[{{van den Bosch} \& {de Zeeuw}(2010)}]{2010MNRAS.401.1770V}
{van den Bosch} R.~C.~E., {de Zeeuw} P.~T., 2010, \mnras, 401, 1770

\bibitem[{{van den Bosch} {et~al.}(2016){van den Bosch}, {Greene}, {Braatz},
  {Constantin}, \& {Kuo}}]{2016ApJ...819...11V}
{van den Bosch} R.~C.~E., {Greene} J.~E., {Braatz} J.~A., {Constantin} A.,
  {Kuo} C.-Y., 2016, \apj, 819, 11

\bibitem[{{van den Bosch} \& {van de Ven}(2009)}]{remco09}
{van den Bosch} R.~C.~E., {van de Ven} G., 2009, \mnras, 398, 1117

\bibitem[{{van den Bosch} {et~al.}(2008){van den Bosch}, {van de Ven},
  {Verolme}, {Cappellari}, \& {de Zeeuw}}]{remco08}
{van den Bosch} R.~C.~E., {van de Ven} G., {Verolme} E.~K., {Cappellari} M.,
  {de Zeeuw} P.~T., 2008, \mnras, 385, 647

\bibitem[{{van der Marel} {et~al.}(1998){van der Marel}, {Cretton}, {de Zeeuw},
  \& {Rix}}]{1998ApJ...493..613V}
{van der Marel} R.~P., {Cretton} N., {de Zeeuw} P.~T., {Rix} H.-W., 1998, \apj,
  493, 613

\bibitem[{{Vasiliev} \& {Zelnikov}(2008)}]{2008PhRvD..78h3506V}
{Vasiliev} E., {Zelnikov} M., 2008, \prd, 78, 083506

\bibitem[{{Wallace} \& {Hinkle}(1996)}]{wallace}
{Wallace} L., {Hinkle} K., 1996, \apjs, 107, 312

\bibitem[{{Watkins} {et~al.}(2013){Watkins}, {van de Ven}, {den Brok}, \& {van
  den Bosch}}]{watkins13}
{Watkins} L.~L., {van de Ven} G., {den Brok} M., {van den Bosch} R.~C.~E.,
  2013, \mnras, 436, 2598

\bibitem[{{Wegg} \& {Gerhard}(2013)}]{wegg13}
{Wegg} C., {Gerhard} O., 2013, \mnras, 435, 1874

\bibitem[{{Zhao}(1996)}]{1996MNRAS.283..149Z}
{Zhao} H., 1996, \mnras, 283, 149

\end{thebibliography}
}




%


\bsp	
\label{lastpage}
\end{document}